\documentclass[aps,pre,reprint,superscriptaddress,amsmath,amssymb]{revtex4-2}
\usepackage{graphicx}
\graphicspath{{figures/}{../figures/}{./}{../}}
\usepackage{subcaption}
\usepackage{bm}
\makeatletter
\long\def\@makecaption#1#2{%
  \par\vskip\abovecaptionskip
  \begingroup
   \small\rmfamily
   \leftskip\z@ \rightskip\z@ \parfillskip\@flushglue
   \parindent1.5em
   \let\\\heading@cr
   \indent\@make@capt@title{#1}{#2}\par
  \endgroup
  \vskip\belowcaptionskip}
\makeatother

\newcommand{\abar}{\bar a}
\newcommand{\zstar}{z^{\ast}}
\newcommand{\ainf}{a_{\infty}}
\newcommand{\eps}{\varepsilon}
\DeclareMathOperator{\Var}{Var}
\DeclareMathOperator{\diag}{diag}

\begin{document}

\title{Stabilization of chaotic ensembles by signed coupling}

\author{Unnikrishna Pillai}
\affiliation{Department of Electrical and Computer Engineering,
New York University Tandon School of Engineering, Brooklyn, New York 11201, USA}

\date{\today}

\begin{abstract}
A globally coupled ensemble of logistic units with parameters above the period-doubling
accumulation point, carrying a signed long-short
weight book, relaxes to a stable macroscopic rest that purely positive coupling through the map outputs cannot
produce at or near full coupling. This rest is a designable single-pole filter: the collective coordinate obeys a
first-order recursion whose pole is set by the book's effective parameter, and whose
small-noise rejection follows in closed form from the squared norm of the book and its
parameter-weighted counterparts, the pole gain, and a memory factor. The sign of the pole grades the rejection by the color of the noise,
monotone and oscillatory damping suppressing anti-persistent and persistent noise and
crossing at white noise. For this output coupling at or near full coupling, signed weights and heterogeneity are
jointly necessary; given heterogeneity the minimum-variance book at each pole is unique and
the variance-optimal pole is interior. Positive coupling through the units' states, with sequential updating or with random
delays can settle identical maps, but only at the single-map fixed point, whose level no choice of
weights can move. At full coupling and for a growing randomized ensemble, a diversified signed book settles the aggregate
from a basin whose guaranteed lower bound approaches full measure, at a rate that rises with
the spread of the parameters. The mechanism holds for any affine family and is
supported numerically beyond it. The construction is, term for term, a constrained
minimum-variance portfolio, so the impossibility is the financial fact that a long-only
portfolio cannot cancel a uniformly positive common factor, illustrated on real equity
data.
\end{abstract}

\maketitle

\paragraph{Introduction.}
Ensembles of coupled chaotic units organize into collective states that no single
unit displays: clustering and phase order, and a non-trivial collective rhythm in
which a macroscopic average oscillates while every element stays chaotic
\cite{kaneko1990,chate1992,kaneko1992}. A complementary line of work, \emph{amplitude
death}, shows that coupling, particularly mixed attractive/repulsive or delayed
coupling, can instead quench oscillations and stabilize a fixed point of coupled
limit-cycle or chaotic flows \cite{ad_review,prasad2005,karnatak2007,resmi2011,sharma2012,repulsive2014}. Microscopic
noise is known to reshape such collective states: it can suppress the effective
dimensionality of the mean field and, for globally coupled logistic maps, enters the
self-consistent mean-field map as a shift of the nonlinearity parameter
\cite{shibata1999,demonte2004}.

Positive coupling can also bring identical chaotic maps to rest, at the unstable fixed
point of the single map. Sinha showed this for randomly rewired coupling through the
neighbors' states rather than their map outputs \cite{sinha2002}, Atmanspacher and Scheingraber for lattices
whose sites are updated in random sequence \cite{atmanspacher2005}, and Masoller and
Mart\'i for coupling with random delays \cite{masoller2005,marti2005}. In each case the
weights are convex, the units are identical, and the rest sits at the fixed point of the
single map, so its level is set by the map whatever the weights. Couplings of both signs
are known to hold quiet states as well: in random networks of sigmoid units the quiescent
state stays stable while the variance of the couplings is below a threshold
\cite{sompolinsky1988}. Those units are not chaotic alone, and the quiet state sits at
zero. Under random global coupling of chaotic maps the mean field can settle near a fixed
value while every unit stays chaotic \cite{alvarez2007}, but the field then fluctuates at
any finite ensemble size and no unit comes to rest. The present work differs from all of
these, since every unit is chaotic on its own and the units differ, the coupling acts
through the map outputs at the same step, where positive weights cannot produce rest at
or near full coupling, and the
weights are designed rather than random, so that a signed book sets both the rest level
and the pole of the collective dynamics.

Here a control-theoretic view is taken of a mechanism for \emph{collective rest} in
discrete ensembles: a signed weight book over an affine family of maps tunes an
\emph{effective parameter}, and when that parameter is driven into the period-one
window the macroscopic average stops moving.
At full coupling a diversified signed book settles an ensemble of units above the
period-doubling accumulation point $\ainf\approx3.5699$, and a lower bound on the stabilized
basin fraction approaches full measure exponentially in the ensemble size along a sequence
of growing randomized ensembles, at a rate that rises with the parameter spread, so
heterogeneity acts as a stabilizing resource. The mechanism holds for any affine
(parameter-scaling) family $g_a=a\,h(x)$ and is supported numerically beyond it, where
the aggregate is no longer a member of the family. This rest is not merely a stable state
but a \emph{designable one-pole filter}, whose single pole is fixed by the book and whose
response to noise, white or long-memory, is given in closed form. The defining feature of
the filter, the freedom to place its pole, is denied to positive (convex) output
coupling, so the resulting Hurst-graded noise rejection is something only signed coupling
can realize there. Under output coupling at and near full coupling, two ingredients are jointly necessary: signed weights (to escape the convex
hull of the unit parameters) \emph{and} heterogeneity of the units (to have a hull to escape).
Heterogeneity is precisely pole-controllability: it gives the parameter vector a second
steering direction independent of the budget, and when it holds the minimum-variance book
at each pole is unique and the variance-optimal pole is an interior minimizer, numerically
unimodal for the lists tested; identical units lock the pole and admit no rest at any
leverage. Although the filtering algebra used here is classical
\cite{hosking1981,beran}, the results here are the signed book that escapes the convex
hull of the parameters under output coupling and the pole and noise design built on it.

\paragraph{Model and rest.}
Let unit $i=1,\dots,N$ carry a logistic map $g_i(x)=a_i x(1-x)$ and couple the ensemble
through a weighted mean field,
\begin{equation}
  x^{(i)}_{n+1}=(1-\eps)\,g_i\!\bigl(x^{(i)}_n\bigr)+\eps\,\bar g_n+\xi^{(i)}_n,
  \quad \bar g_n=\sum_{k}w_k\,g_k\!\bigl(x^{(k)}_n\bigr),
  \label{eq:model}
\end{equation}
with $\sum_k w_k=1$ and weights of either sign; $w_k>0$ is a long (diffusive)
position, $w_k<0$ a short (inhibitory) one. The units couple through their map outputs
at the same step, the form of globally coupled maps \cite{kaneko1990b}, called output
coupling below. The per-unit noise $\xi^{(i)}_n$ is
zero-mean; in general it carries a cross-unit covariance
$C_{ik}=\langle\xi^{(i)}_n\xi^{(k)}_n\rangle$ and a separable common temporal
autocovariance $\rho(m)$ ($\rho(0)=1$), i.e.\ $\langle\xi^{(i)}_n\xi^{(k)}_{n+m}\rangle
=C_{ik}\,\rho(m)$. The closed-form results below take the units \emph{independent},
$C=\sigma^2 I$ (variance $\sigma^2$, white when $\rho(m)=0$ for $m\ne0$); independent colored
noise follows by carrying $\rho$, and temporally white cross-correlated noise by carrying
$C$, the case realized empirically. The collective coordinate and the effective parameter are
\begin{equation}
  z_n=\sum_i w_i x^{(i)}_n,\qquad \abar=\sum_i w_i a_i.
\end{equation}
At full coupling every unit is driven by the same field,
$x^{(i)}_{n+1}=\bar g_n+\xi^{(i)}_n$ with $\bar g_n=\sum_k w_k g_k(x^{(k)}_n)$, so after
one step the units share the common value $\bar g_n$ and differ only by their own noise.
The deviation of each unit from the collective coordinate is then
$x^{(i)}_n-z_n=\xi^{(i)}_{n-1}-\bar\xi_{n-1}$, with $\bar\xi_n=\sum_i w_i\xi^{(i)}_n$, of the order of the noise and not growing:
the ensemble synchronizes. On the noise-free synchronized set all units equal $z_n$,
the field collapses to $\sum_k w_k g_k(z_n)=\abar\,z_n(1-z_n)$, so there the collective
coordinate obeys the scalar logistic map $z_{n+1}=\abar\,z_n(1-z_n)$ exactly. Restoring
the noise, the collective coordinate follows the reduced map
\begin{equation}
  z_{n+1}=\abar\,z_n(1-z_n)+\bar\xi_n,
  \label{eq:eff}
\end{equation}
which keeps the common-mode forcing but not the further noise-dispersion terms that the
noise-driven spread of the units generates, so it is exact only on the noise-free manifold. The
deterministic skeleton, the fixed point, the stability window and the barrier, lives on
the synchronized set. The exact noisy collective equation, and the one-pole recursion
used for the noise analysis as its first-order small-noise linearization, are derived
below.

The noise-free map has the period-one fixed point $\zstar=1-1/\abar$, stable for
$\abar\in(1,3)$. The barrier that underlies everything here, and the reason positive
coupling is excluded, is the following elementary fact about convex averages.

\begin{quote}\emph{Proposition (no long-only rest).} Under the output coupling of
Eq.~\eqref{eq:model}, if $w_i\ge0$ for all $i$, $\sum_i w_i=1$, and every unit lies above
the period-doubling accumulation point ($\ainf<a_i\le4$ \cite{feigenbaum1978}), then $\abar=\sum_i w_i a_i\in
[\min_i a_i,\max_i a_i]\subset(\ainf,4]$, so $\abar>3$ and the nonzero fixed point
$\zstar=1-1/\abar$ is unstable (its multiplier $2-\abar$ has $|2-\abar|>1$), so no stable
period-one collective rest exists. Reaching the rest window $\abar\in(1,3)$ forces
$\sum_i w_i a_i<\min_i a_i$, hence at least one $w_i<0$. \emph{(Proof: a convex
combination lies between its smallest and largest entries, and $\min_i a_i>\ainf>3$.\
$\square$)}\end{quote}

\noindent Convexity pins the effective drive inside the hull of the $a_i$; only
subtraction escapes it. The asymmetry of the logistic family about $\ainf$ makes the
barrier sharp: below $\ainf$ the map is non-chaotic on essentially the whole interval
(attracting fixed points for $1<\abar<3$, then stable $2^k$-cycles up to $\ainf$),
whereas above $\ainf$ chaos has positive measure \cite{jakobson1981} and the
non-chaotic set is the thin scatter of periodic windows. Numerically, the fraction
with non-positive Lyapunov exponent is $100\%$ on $(2.5,\ainf)$ but only ${\sim}10\%$
on $(\ainf,4)$. A long-only book therefore reaches at best a periodic window, never a fixed point;
genuine fixed-point rest in $1<\abar<3$ requires a short.
On one ensemble of forty units above the accumulation point, convex weights leave the mean
field chaotic while a signed book on the \emph{same} units rests.
The Proposition is stated on the synchronized set reached at full coupling.
Appendix~\ref{app:below} extends the exclusion to open windows of coupling near zero and
near one, uniformly over convex books, and between these windows it is supported
numerically for two units.

\emph{Remark (other positive couplings).} The Proposition concerns output coupling, and for
identical units, $a_i\equiv a>3$, this coupling excludes a homogeneous rest at every
coupling and for any weights summing to one, signed or not. At the homogeneous fixed point
$x^{(i)}=1-1/a$ the Jacobian is $(2-a)M$ with $M=(1-\eps)I+\eps\,\mathbf 1 w^\top$, where $I$ is
the identity, $\mathbf 1$ the all-ones vector and $w$ the vector of weights, and
$M\mathbf 1=\mathbf 1$, so the uniform mode carries the multiplier $2-a$, of modulus above
one. Positive coupling of other forms does rest identical units at $1-1/a$. Coupling
through the neighbors' states \cite{sinha2002} gives the uniform mode the multiplier
$(1-\eps)(2-a)+\eps$, which enters the unit circle at strong enough coupling, and random
links bring the remaining modes inside as well. Random sequential updating \cite{atmanspacher2005} feeds already-updated states into the
coupling, and random delays \cite{masoller2005,marti2005} feed past states. Each of these
mechanisms puts into the average something other than the current map outputs, namely the
states themselves, already-updated values or past states, so the field is no longer a
convex average of chaotic maps at the current step, and the hull bound on $\abar$
that underlies the Proposition does not apply. In all of them the units are identical and
the weights convex, so neither a short nor heterogeneity is necessary there, but the rest
sits at $1-1/a$, fixed by the map whatever the weights. The signed book of
Eq.~\eqref{eq:model} instead sets both the rest level $\zstar=1-1/\abar$ and the pole of
the linearized dynamics.

\begin{figure}[htp]
  \centering
  \begin{subfigure}{0.48\linewidth}
    \centering
    \includegraphics[width=\linewidth]{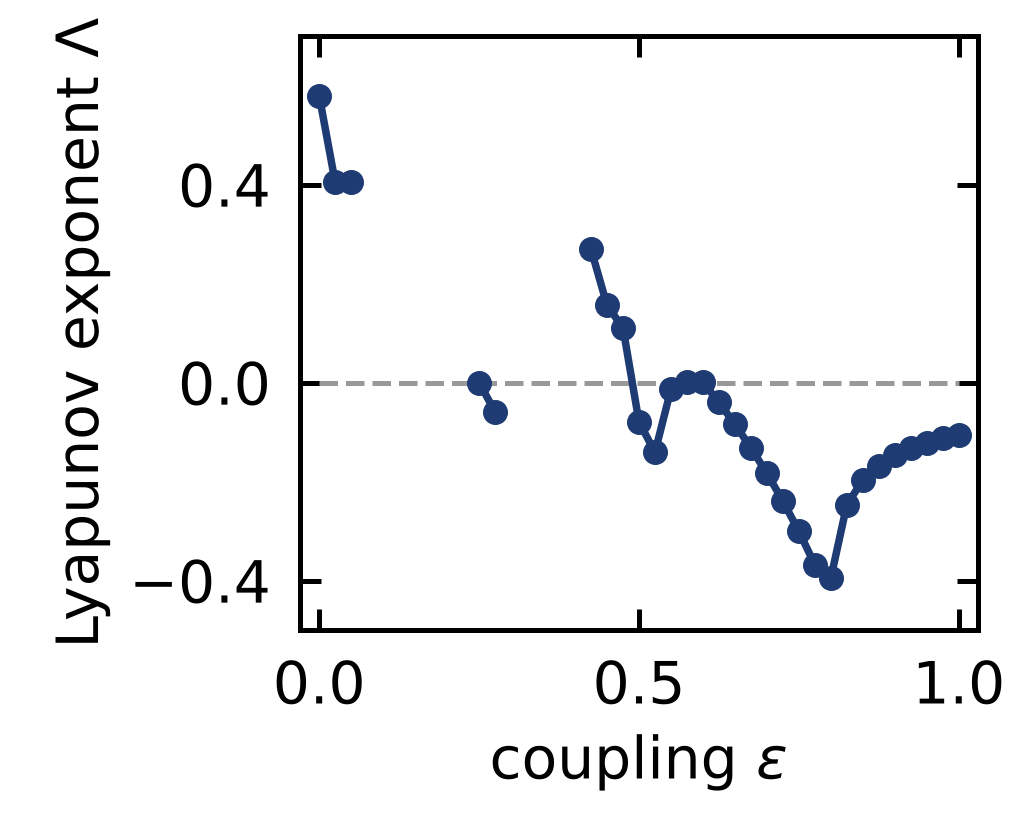}
    \caption{}
    \label{fig1a}
  \end{subfigure}
  \hfill
  \begin{subfigure}{0.48\linewidth}
    \centering
    \includegraphics[width=\linewidth]{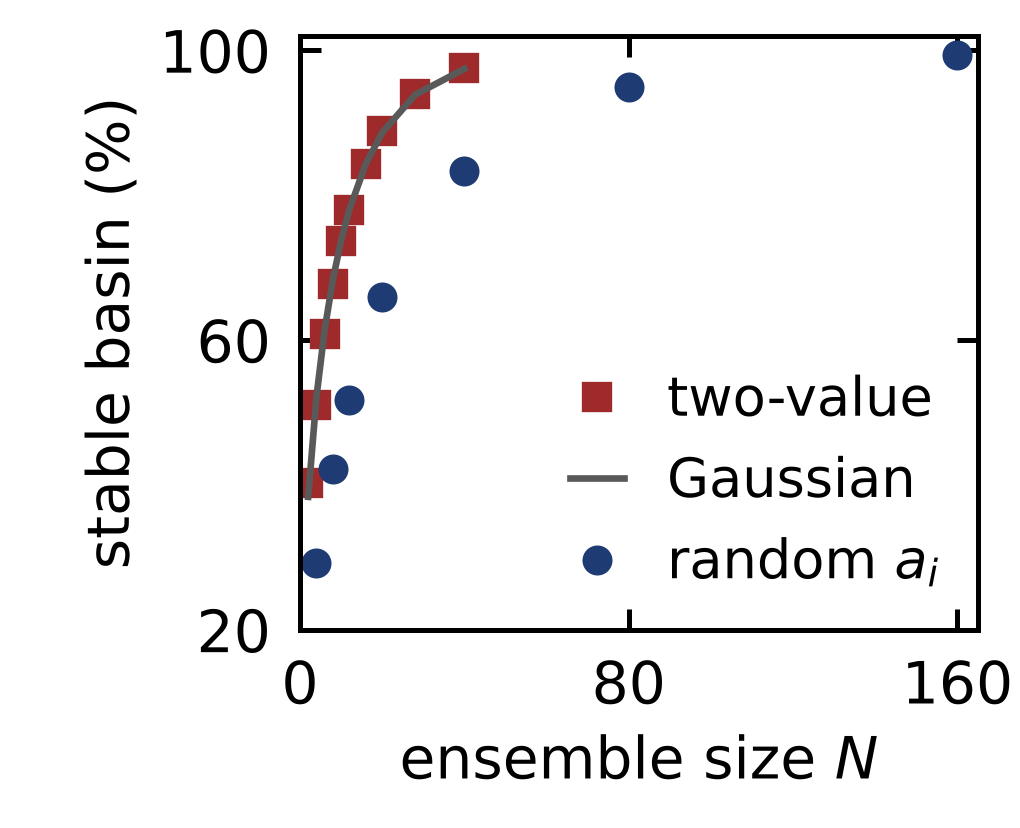}
    \caption{}
    \label{fig1b}
  \end{subfigure}
  \caption{Reliability of rest (simulation). (a) Largest Lyapunov exponent $\Lambda$ of
  the full two-unit map versus coupling $\varepsilon$ for chaotic units $a_1=3.60$,
  $a_2=3.95$ with signed weights $w=(3,-2)$, $\abar=2.9$: chaotic when uncoupled
  ($\Lambda=+0.58$) and at rest for $\varepsilon\gtrsim0.6$; the negative values at weaker
  coupling belong to periodic orbits, and missing points are couplings at which the
  trajectory escapes. (b) Stable basin versus ensemble size $N$ at $\varepsilon=1$.
  Squares: the two-value book with weights $6/N$ and $-4/N$ ($\abar=2.9$, $L=5$), with its
  Gaussian concentration estimate (line, closed form). Circles: heterogeneous units
  $a_i\sim\mathrm{Unif}(\ainf,4)$ stabilized by the minimum-norm book at $\abar_t=2.8$.
  Both approach full measure as $N$ grows.}
  \label{fig:robust}
\end{figure}

\paragraph{Reliability of rest: a robustness theorem.}
The impossibility makes a short \emph{necessary}; the next result bounds from below the basin
of a \emph{diversified} short, and the bound approaches one as the (randomized) ensemble
grows. Work at full slaving
$\varepsilon=1$, where the units collapse
to the mean field in one step: with
$z_1=\sum_i w_i a_i\,x_0^{(i)}\!\bigl(1-x_0^{(i)}\bigr)$ the aggregate thereafter obeys
the reduced map, and for $\abar\in(1,3)$ it settles to $\zstar$ iff the first iterate
lands in the unit interval, $z_1\in(0,1)$. For initial states drawn uniformly on
$[0,1]^N$ the increments $x_0^{(i)}(1-x_0^{(i)})$ are i.i.d.\ on $[0,\tfrac14]$ with mean
$\tfrac16$ and variance $\tfrac1{180}$, so $\mathbb E[z_1]=\abar/6\in(0,\tfrac12)$, an
interior target. Each summand occupies an interval of width $|w_i|a_i/4$, and
Hoeffding's inequality bounds the failure probability,
\begin{equation}
  \Pr\bigl[z_1\notin(0,1)\bigr]\le
  2\exp\!\left(-\frac{32(\abar/6)^2}{\sum_i w_i^2 a_i^2}\right).
  \label{eq:hoeffding}
\end{equation}
(Independence is used only for the initial draw; the dynamically generated correlations
that can defeat the law of large numbers in a sustained chaotic regime \cite{kaneko1990}
do not enter this one-step criterion.) Since $\sum_i w_i^2 a_i^2\le a_+^2\|w\|_2^2$
with $a_+=\max_i a_i$ and, by Cauchy--Schwarz at fixed leverage $L\equiv\|w\|_1$, $\|w\|_2^2\ge L^2/N$ with
equality for an evenly spread book, an evenly spread fixed-leverage signed allocation
has stable basin
\begin{equation}
  \mathcal B(N)\ \ge\ 1-2\,e^{-\gamma N/L^2}\ \xrightarrow[N\to\infty]{}\ 1,
  \qquad \gamma=\frac{32(\abar/6)^2}{a_+^2},
  \label{eq:basin}
\end{equation}
so for a diversified long-short book of arbitrarily many units above the accumulation point
the guaranteed fraction of randomized initial conditions that settle tends to one as $N$
grows along this sequence; concentrating the book raises $\|w\|_2^2$ and weakens the
guarantee, so even spreading optimizes the bound. Diversification does not create the stability (the shorts
do), but it makes the stability insensitive to where the ensemble begins.
Figure~\ref{fig:robust}(b) shows the measured basin of the two-value book with weight $6/N$
on half the units, at $a=3.60$, and $-4/N$ on the other half, at $a=3.95$ ($L=5$,
$\abar=2.9$), rising from $40\%$ at $N=2$ to $97.5\%$ at $N=40$. At these sizes
the Hoeffding bound is loose, and the measured basin follows the sharper Gaussian estimate.

The rest found at $\varepsilon=1$ continues to an open coupling window. There it is the
slaved diagonal fixed point, and by the implicit function theorem a nearby fixed point of
the full system Eq.~\eqref{eq:model} persists and stays locally attracting as
$\varepsilon$ decreases from $1$ (the transverse multipliers, $O(1-\varepsilon)$
there, remain inside the unit circle), so rest survives on an interval
$\varepsilon\in[\varepsilon^\ast,1]$; the theorem carries the fixed point and its local
attraction, not the basin measure. Figure~\ref{fig:robust}(a) shows the largest Lyapunov
exponent $\Lambda$ of the full two-unit map ($a_1=3.60$, $a_2=3.95$, both chaotic;
$w=(3,-2)$, $\abar=2.9$), which is $+0.58$ uncoupled, the exponent of the $a_2=3.95$ unit,
and $-0.39$ at $\varepsilon=0.8$. For this book the rest is a stable fixed point for
$\varepsilon\gtrsim0.6$, where a complex pair of multipliers crosses the unit circle, and
the negative values of $\Lambda$ at weaker coupling belong to periodic orbits, not to rest.
Appendix~\ref{app:below} proves the continuation of rest below full coupling and shows
that no book of bounded leverage rests at weak coupling.

\emph{Heterogeneity sets the rate.} For units with empirical mean $m_a$ and spread
$q=N^{-1}\sum_i(a_i-m_a)^2$, the minimum-norm book reaching a target
$\abar_t\in(1,3)$ is the affine book
$w_i=N^{-1}[1+(\abar_t-m_a)(a_i-m_a)/q]$ (derived in Appendix~\ref{app:opt}), each unit
shorted in proportion to its parameter excess $a_i-m_a$, with squared norm
$\|w\|_2^2=N^{-1}[1+(m_a-\abar_t)^2/q]$. Since $\|w\|_2^2$ controls the bound~\eqref{eq:basin}, the
guarantee factorizes,
\begin{equation}
  \mathcal B(N)\ge1-2e^{-\tilde\gamma N},\qquad
  \tilde\gamma=\gamma\;\frac{q}{\,q+(m_a-\abar_t)^2\,},
  \label{eq:hetrate}
\end{equation}
an edge prefactor $\gamma$ times a \emph{diversity factor} $q/[q+(m_a-\abar_t)^2]$ that,
at fixed $N$, mean $m_a$, target $\abar_t$ and bound $a_+$ (so that $\gamma$ is held
fixed), is strictly increasing in the spread,
$\partial\tilde\gamma/\partial q=\gamma\,(m_a-\abar_t)^2/[q+(m_a-\abar_t)^2]^2>0$.
Figure~\ref{fig:robust}(b) also shows (circles) heterogeneous units drawn uniformly above
$\ainf$ and hedged by this book at $\abar_t=2.8$, whose basin likewise approaches full
measure as $N$ grows. Heterogeneity is known to induce order in globally coupled chaotic maps
\cite{shibata1997}; here it enters the exponent of the bound as a \emph{resource}, and the
greater the parameter spread, the faster the guaranteed basin approaches one. As $q\to0$ the rate vanishes and
no finite-leverage hedge exists; an identical ensemble cannot be reweighted to rest,
since $\abar=a_1$ is locked for every book. This is the dynamical face of the same
heterogeneity that reappears below as \emph{pole-controllability}: there it makes the
rest pole placeable and the optimal book unique, here it raises the guaranteed basin,
both scaling with the spread $q$.

\begin{figure}[htp]
  \centering
  \includegraphics[width=0.74\columnwidth]{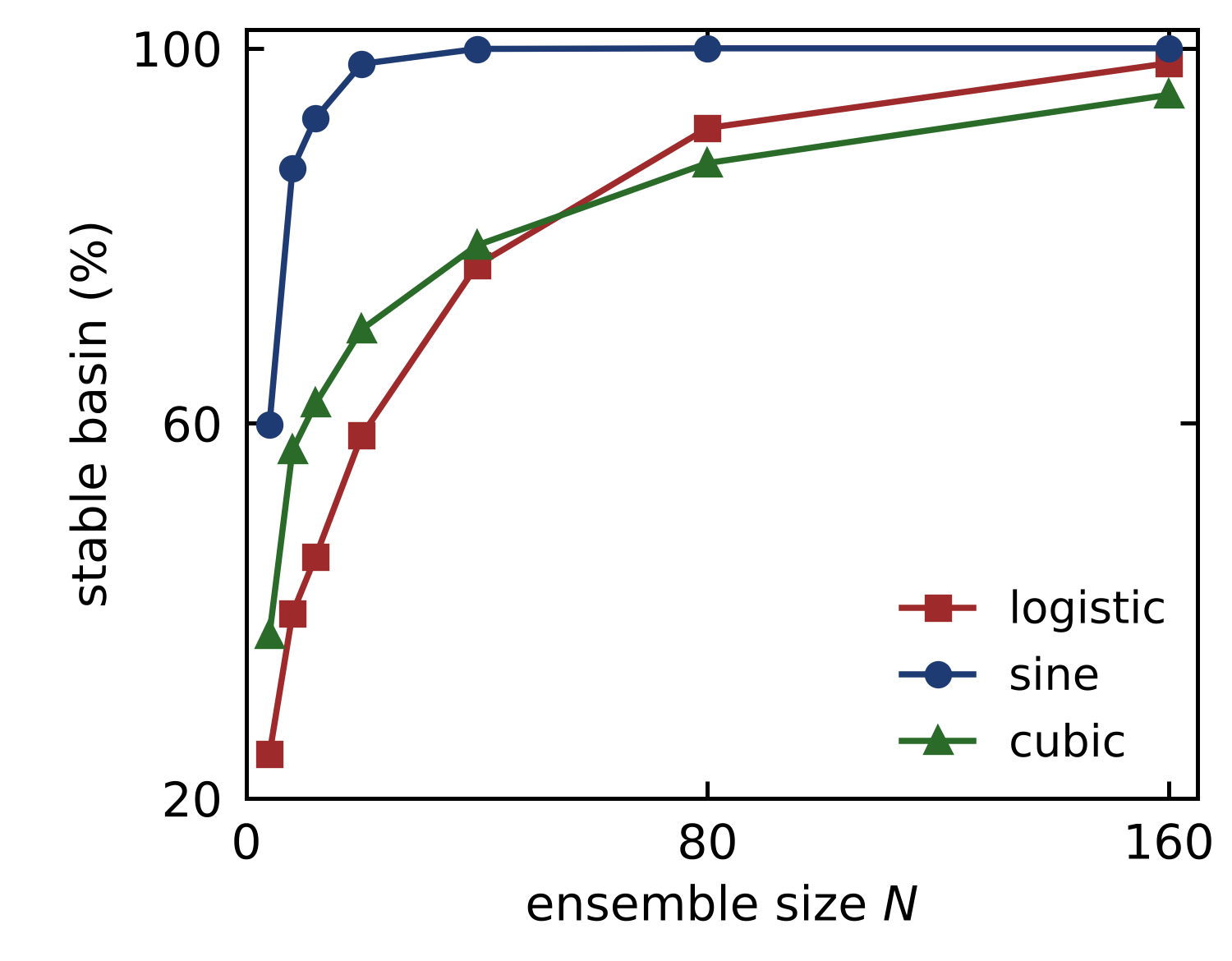}
  \caption{Other one-parameter families $g_a=a\,h(x)$ (simulation). Stable basin
  versus ensemble size $N$ at $\varepsilon=1$ for heterogeneous logistic, sine, and
  cubic ensembles, with $a_i$ drawn uniformly from $(3.60,3.98)$, $(0.90,1.00)$ and
  $(0.80,1.00)$, above each family's chaos threshold, and hedged by the
  minimum-norm book to a value in its own rest band ($\abar=2.8$ logistic, $0.45$ sine,
  $0.5$ cubic). For the odd sine shape the count
  includes the mirror rest $-\zstar$. All approach full measure as $N$ grows.}
  \label{fig:univ}
\end{figure}

\paragraph{Generality: one-parameter families and a non-affine test.}
The construction uses only that each unit has the form $g_a(x)=a\,h(x)$, a parameter
scaling a fixed unimodal shape $h$. Then $\sum_k w_k g_{a_k}(x)=\abar\,h(x)$ is again a
member of the family, so the synchronized reduction, the impossibility Proposition and
the minimum-norm book carry over unchanged. A concentration bound analogous to
Eqs.~\eqref{eq:hoeffding}--\eqref{eq:hetrate} also follows whenever $h$ is bounded on $[0,1]$,
but its constants and its attraction criterion depend on the range, mean, and basin
geometry of $h$. The values $\tfrac16$, $\tfrac1{180}$, $[0,\tfrac14]$ and
$z_1\in(0,1)$ used above are specific to $h(x)=x(1-x)$ and must be recomputed for each
family, with $\ainf$ and the rest band likewise replaced by those of $h$.
Figure~\ref{fig:univ} illustrates this for heterogeneous logistic [$h=x(1-x)$], sine
[$h=\sin\pi x$], and cubic [$h=\tfrac{27}{4}x(1-x)^2$] ensembles, each drawn above its
own threshold and hedged to its own rest band; in each the fraction of initial conditions
that settle approaches one as $N$ grows. The sine shape is odd, so the mirror point
$-\zstar$ is an equally stable rest of the aggregate, and the count for that family
includes it.

\begin{figure}[htp]
  \centering
  \begin{subfigure}{0.48\linewidth}
    \centering
    \includegraphics[width=\linewidth]{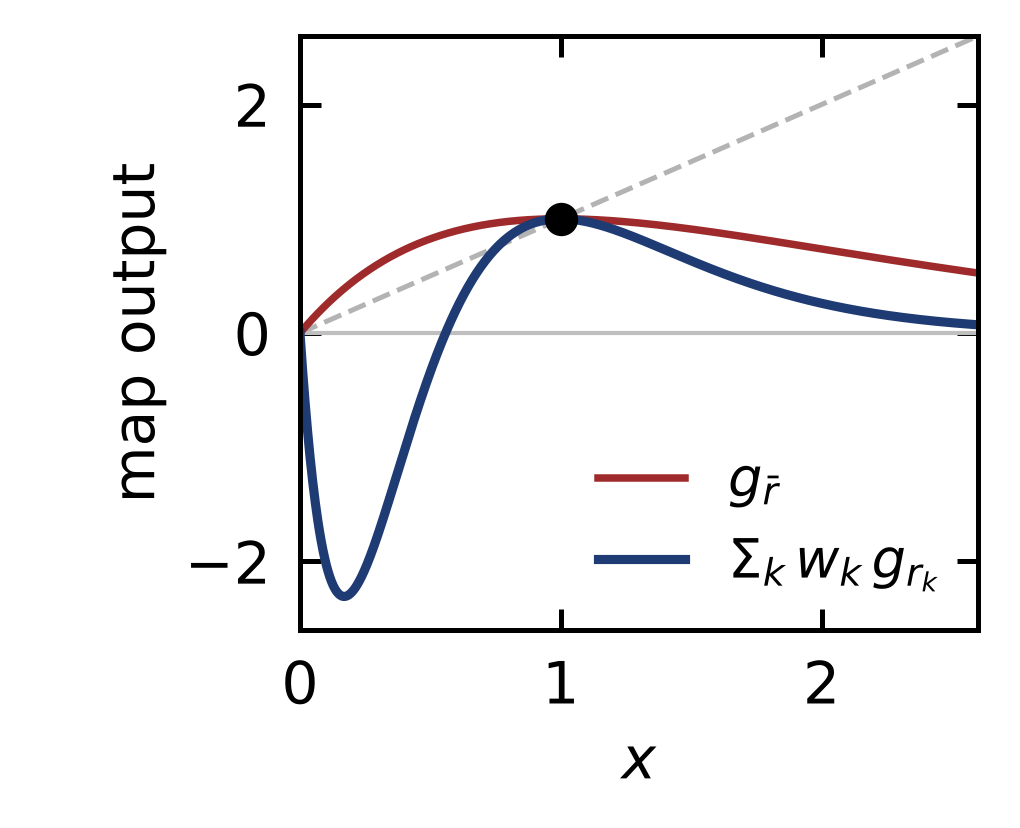}
    \caption{}
    \label{fig3a}
  \end{subfigure}
  \hfill
  \begin{subfigure}{0.48\linewidth}
    \centering
    \includegraphics[width=\linewidth]{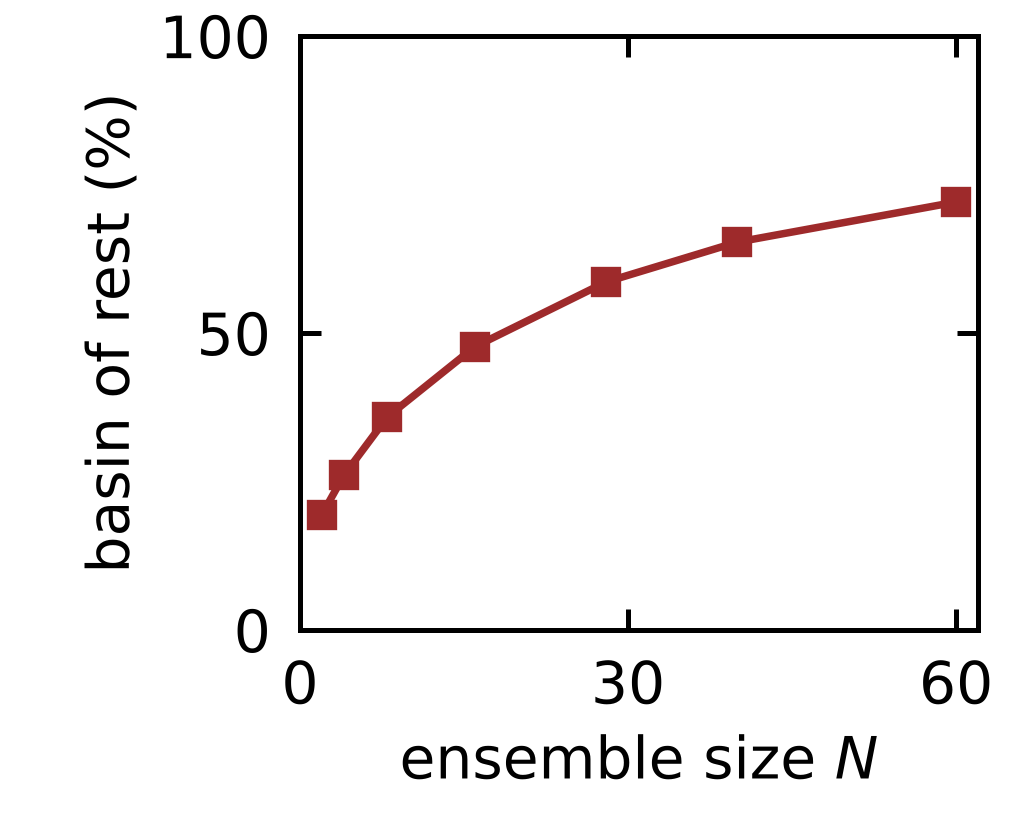}
    \caption{}
    \label{fig3b}
  \end{subfigure}
  \caption{Non-affine control (Ricker map $g_r(x)=x\,e^{r(1-x)}$, $r$ in the
  exponent). (a) Closed-form map curves, in which the signed-book aggregate $\sum_k w_k g_{r_k}$
  (blue), negative for $x\lesssim0.56$ and unimodal on its positive branch, does not
  coincide with the single Ricker map $g_{\bar r}$ (red); by linear independence of the
  exponentials it is generally not a member of the Ricker family. The diagonal and the
  parameter-free rest $z=1$ (dot) are shown. (b) Basin of the collective rest
  versus ensemble size $N$ at $\varepsilon=1$ (simulation), measured on a bounded operating
  region about $z=1$, for a two-value signed book ($r_A=2.80$, $r_B=3.80$, $\bar r=1$).}
  \label{fig:ricker}
\end{figure}

These families share the affine form, for which the aggregate collapses to a single
effective parameter; the sharper question is whether the mechanism survives when it does
not. \emph{A non-affine test.} For the Ricker map $g_r(x)=x\,e^{r(1-x)}$ the parameter
sits in the exponent, so for distinct $r_k$ with nonzero weights the signed-book
aggregate $\sum_k w_k g_{r_k}(x)=x\sum_k w_k e^{r_k(1-x)}$ is generally not a member of
the one-parameter Ricker family: linear independence of the exponentials prevents it
from reducing to $x\,e^{r(1-x)}$ except in degenerate cases [Fig.~\ref{fig:ricker}(a)].
The two values used here are nonetheless individually chaotic: the single-map Lyapunov exponents are $+0.24$ at $r_A=2.80$ and $+0.55$ at
$r_B=3.80$, both above the chaos onset $r_c\approx2.69$ and away from the periodic
windows that also occur there. A convex book leaves the aggregate chaotic, and a signed
book drives the ensemble to the \emph{parameter-free} rest $z=1$, a fixed point of the
aggregate for any book with $\sum_k w_k=1$, multiplier $1-\bar r$ with $\bar r=\sum_k w_k r_k$, superstable at
$\bar r=1$, whose basin grows with $N$, from $19\%$ to $72\%$ for a two-value book
[Fig.~\ref{fig:ricker}(b)]. Because the Ricker state space $[0,\infty)$ is unbounded, the basin is
measured on a bounded operating region about $z=1$, and only the closed-form bound
Eq.~\eqref{eq:hoeffding} relies on the unit-interval image of the logistic family; the
stabilization itself is a property of the mean-field/threshold structure, not of any
closed-form parameter average.

\paragraph{Rest as a tunable pole.}
At full coupling the deviation of each unit from the collective coordinate is the demeaned
noise of the previous step, $\delta^{(i)}_n=\xi^{(i)}_{n-1}-\bar\xi_{n-1}$ with
$\sum_i w_i\delta^{(i)}_n=0$. Because each map is quadratic the expansion
$g_k(z+\delta)=a_k[z(1-z)+(1-2z)\delta-\delta^2]$ is exact, and summing the update with
the weights gives the exact dynamics of the collective coordinate,
\begin{equation}
\begin{aligned}
  z_{n+1}&=\abar\,z_n(1-z_n)+\bar\xi_n+(1-2z_n)\,S_n-Q_n,\\
  S_n&=\sum_k w_k a_k\,\delta^{(k)}_n,\qquad
  Q_n=\sum_k w_k a_k\,(\delta^{(k)}_n)^2 .
\end{aligned}
  \label{eq:exactz}
\end{equation}
Equation~\eqref{eq:eff} keeps the first two terms, while the term $S_n$ is linear in the
noise, of the same order as $\bar\xi_n$, and $Q_n$ is quadratic. Writing $u_n=z_n-\zstar$
and linearizing about $\zstar$, where the slope is $\lambda=2-\abar$ and $1-2\zstar=\lambda/\abar$,
\begin{equation}
  u_{n+1}=\lambda\,u_n+\bar\xi_n+\frac{\lambda}{\abar}\,S_n,\qquad
  \lambda=2-\abar,\quad|\lambda|<1 .
  \label{eq:ar1}
\end{equation}
For this first-order (one-pole) recursion, in the transform variable $\nu$ with
$\nu^{-1}$ a one-step delay, the filter from the forcing to the rest deviation $u$ is
$\Phi(\nu)=1/(1-\lambda\nu^{-1})$, with a single pole at $\nu=\lambda$. The book \emph{places} that pole: since $\lambda=2-\abar$ with
$\abar=\sum_i w_i a_i$, choosing the weights slides $\lambda$ along the real axis in
$(-1,1)$. The setting $\abar=2$ puts it at the origin (flat, deadbeat decay), $\abar<2$ on the
positive axis (low-pass, monotone decay), $\abar>2$ on the negative axis (high-pass,
alternating decay), and $\abar\to1,3$ against the unit circle (marginal, long memory).
Read this way the impossibility above is a statement about pole placement: positive
output coupling cannot reach $\abar\in(1,3)$, hence cannot put the pole inside the unit
circle, so no stable filter exists without subtraction.

The forcing of \eqref{eq:ar1} is $f_n=\bar\xi_n-\lambda\,\bar\xi_{n-1}
+(\lambda/\abar)\,A_{n-1}$, where $A_n=\sum_i w_i a_i\,\xi^{(i)}_n$ is the
parameter-weighted noise and $S_n=A_{n-1}-\abar\,\bar\xi_{n-1}$. The common-mode part
$\bar\xi_n-\lambda\bar\xi_{n-1}$ carries a zero at $\nu=\lambda$ that cancels the pole of
$\Phi$, so the common-mode noise reaches the rest with unit gain and only the
parameter-weighted noise is shaped by the pole. For independent units of per-unit
variance $\sigma^2$ and temporal autocovariance $\rho(m)$, passing $f$ through $\Phi$ and
collecting terms gives, to first order in the noise, the rest variance
$\Var(z_\infty)$, the stationary variance of the collective coordinate as $n\to\infty$,
\begin{equation}
\begin{aligned}
  \Var(z_\infty)={}&\sigma^2\|w\|_2^2
  +\frac{\sigma^2\,(G-1)}{\abar}\sum_i w_i^2 a_i\\
  &+\frac{\sigma^2\,G\,\lambda^2}{\abar^2(1-\lambda^2)}\sum_i w_i^2 a_i^2 ,
\end{aligned}
  \label{eq:main}
\end{equation}
with the noise-color factor
\begin{equation}
  G(\lambda;\rho)=\sum_{m=-\infty}^{\infty}\rho(m)\,\lambda^{|m|}
  =1+2\sum_{m\ge1}\rho(m)\,\lambda^{m}.
  \label{eq:G}
\end{equation}
At fixed weights the common-mode floor $\sigma^2\|w\|_2^2$ has unit transfer gain (no
pole amplification); the parameter-weighted terms carry the pole gain and the color. For
cross-correlated unit noise of covariance $C$ (still temporally white) the floor becomes
$w^\top C\,w$ and the parameter-weighted term becomes
$\kappa\,w^\top\!\diag(\mathbf a)\,C\,\diag(\mathbf a)\,w$ with
$\mathbf a=(a_1,\dots,a_N)$ and $\kappa=\lambda^2/[\abar^2(1-\lambda^2)]$, so the rest variance is the quadratic form
$w^\top[\,C+\kappa\,\diag(\mathbf a)\,C\,\diag(\mathbf a)\,]w$; the parameter-weighted
covariance is $\diag(\mathbf a)\,C\,\diag(\mathbf a)$, not $C$ alone.
Equation~\eqref{eq:main} is the overlap of the forcing spectrum with the one-pole gain,
\begin{equation}
  \Var(z_\infty)=\frac{1}{2\pi}\int_{-\pi}^{\pi}
  \frac{P_f(\omega)}{|1-\lambda e^{-j\omega}|^2}\,d\omega ,
  \label{eq:spectral}
\end{equation}
where $|1-\lambda e^{-j\omega}|^{-2}=(1-2\lambda\cos\omega+\lambda^2)^{-1}$ is the squared
magnitude response of the one-pole filter $\Phi$ and $P_f$ is the spectral density of the
forcing $f$.

\emph{White noise, independent units.} For $\rho(m)=0$ at $m\ne0$, $G=1$, the middle
term drops and
\begin{equation}
\begin{aligned}
  \Var(z_\infty)={}&\sigma^2\|w\|_2^2\\
  &+\frac{\sigma^2\,(2-\abar)^2}{\abar^2(\abar-1)(3-\abar)}\sum_i w_i^2 a_i^2 ,
\end{aligned}
  \label{eq:varbook}
\end{equation}
using $1-\lambda^2=(\abar-1)(3-\abar)$. At fixed weights the common-mode floor
$\sigma^2\|w\|_2^2$ carries no pole gain; the parameter-weighted term is a well over the
rest window that vanishes at the superstable book $\abar=2$ ($\lambda=0$) and diverges at
the edges $\abar\to1,3$. Equations~\eqref{eq:main}--\eqref{eq:varbook} are first-order
small-noise predictions: they keep the noise forcing of \eqref{eq:ar1} and drop the
curvature $-\abar\,u_n^2$, the quadratic noise term $Q_n$, and the products $u_n S_n$,
all of higher order in the noise.

\begin{figure}[htp]
  \centering
  \begin{subfigure}[t]{0.48\linewidth}
    \centering
    \includegraphics[width=\linewidth]{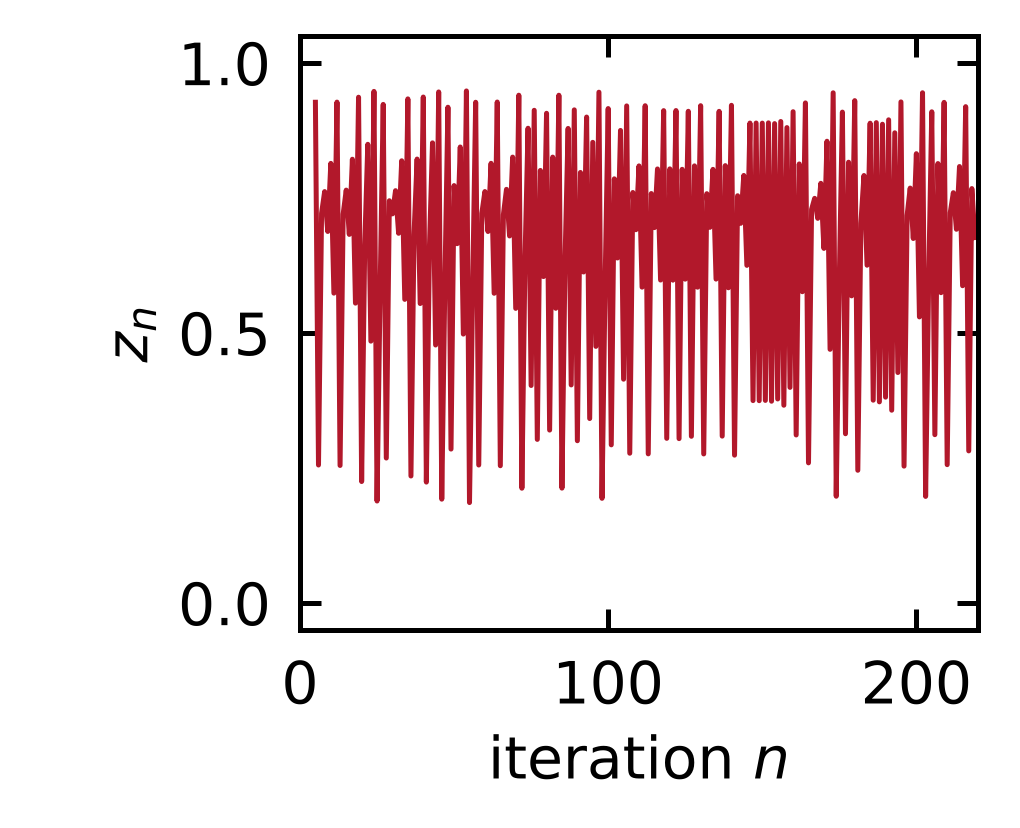}
    \caption{}
    \label{fig4a}
  \end{subfigure}
  \hfill
  \begin{subfigure}[t]{0.48\linewidth}
    \centering
    \includegraphics[width=\linewidth]{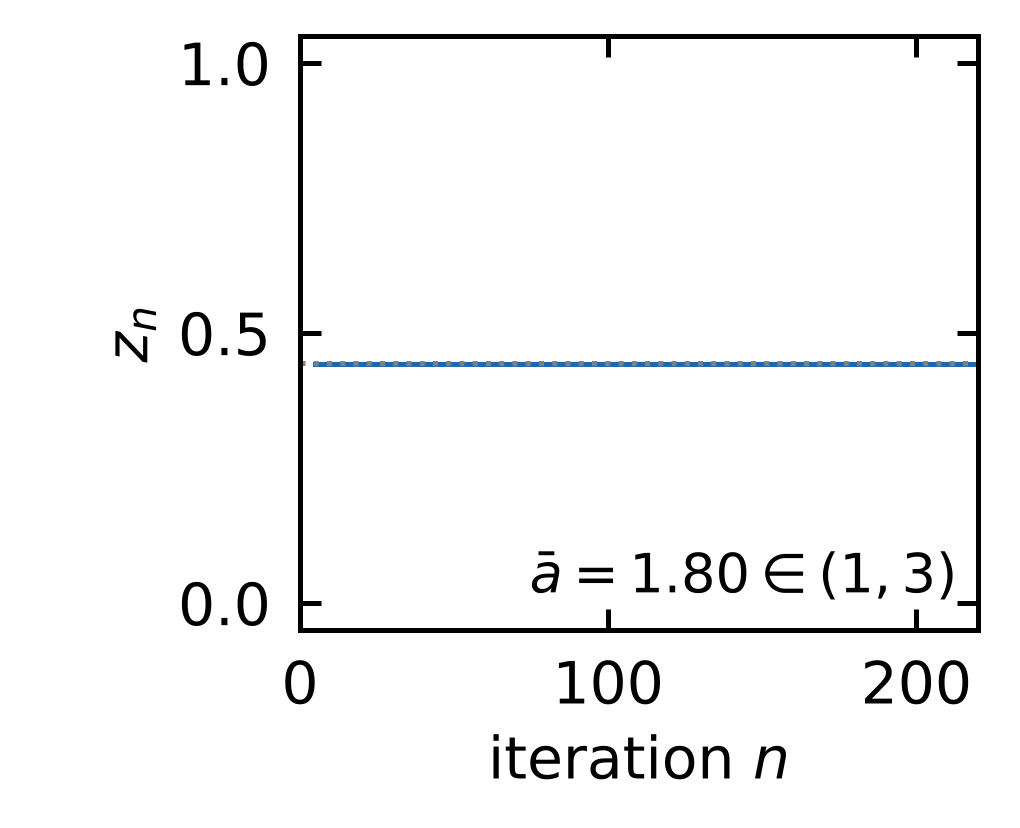}
    \caption{}
    \label{fig4b}
  \end{subfigure}
  \vspace{1em}
  \begin{subfigure}[t]{0.48\linewidth}
    \centering
    \includegraphics[width=\linewidth]{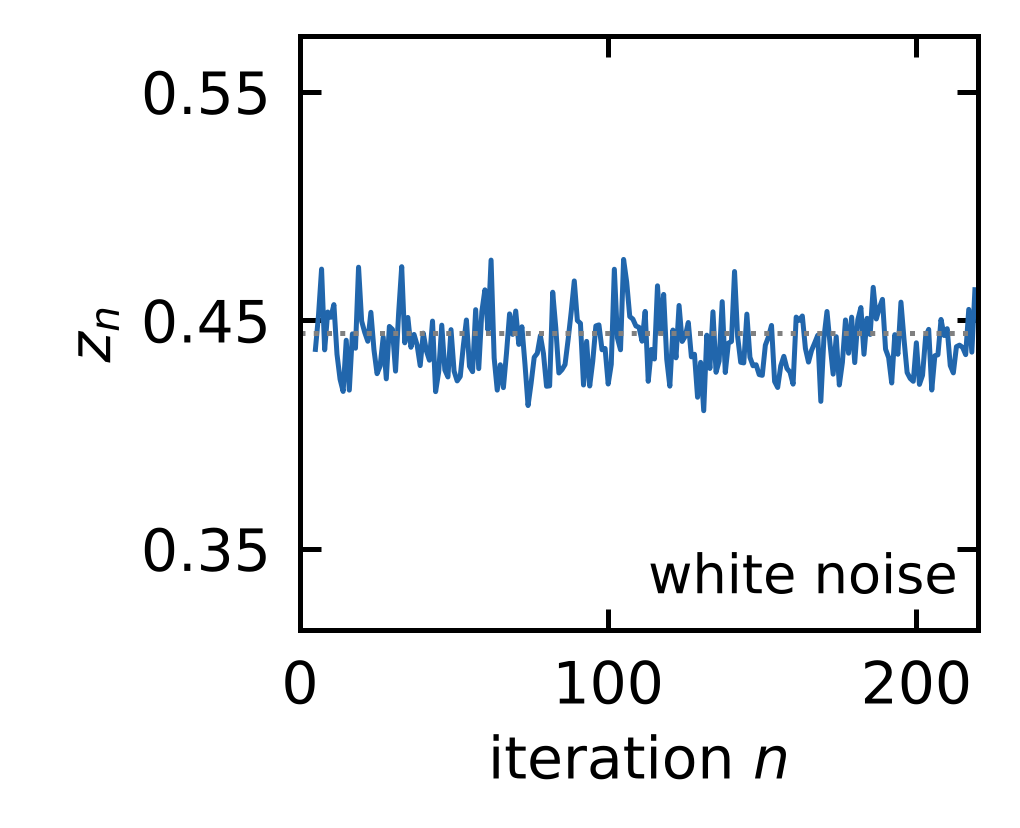}
    \caption{}
    \label{fig4c}
  \end{subfigure}
  \hfill
  \begin{subfigure}[t]{0.48\linewidth}
    \centering
    \includegraphics[width=\linewidth]{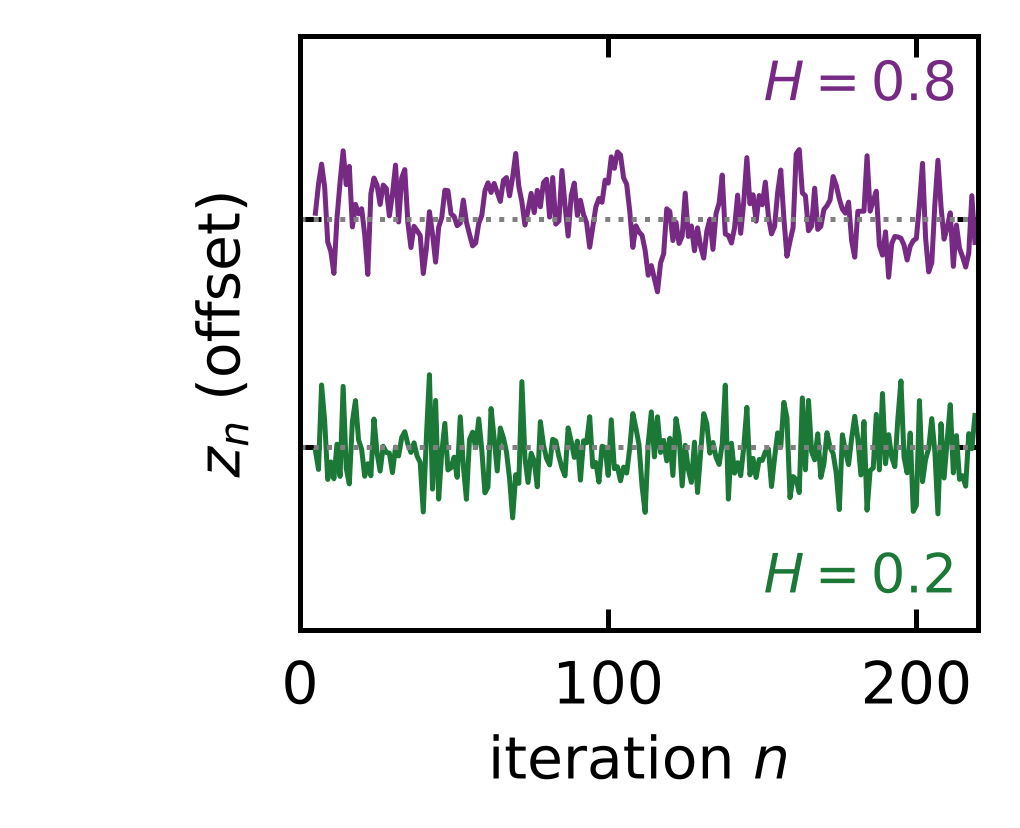}
    \caption{}
    \label{fig4d}
  \end{subfigure}
  \caption{Collective rest under noise (simulation) on one ensemble of $N=40$ logistic
  units, every unit above the accumulation point ($a_i\in(3.60,4.00)$, all $>\ainf$),
  $\eps=0.95$, per-unit noise level $\sigma=0.005$ in (c) and (d).
  (a) Convex (long-only) weights: $\abar=3.79>\ainf$ and the mean field never rests.
  (b) A signed minimum-norm book with $\abar=1.80\in(1,3)$ on the same units: rest near
  $\zstar=1-1/\abar$. (c) The signed book with white per-unit noise: rest persists as a
  tight small-noise cloud over the shown horizon in the bounded (saturated)
  implementation. (d) With
  fractional Gaussian per-unit noise, persistent driving ($H=0.8$) gives slower and
  somewhat wider excursions than anti-persistent driving ($H=0.2$), since this book is
  low-pass ($\lambda=2-\abar>0$); the two traces are offset by $\pm0.05$ for clarity.}
  \label{fig:noise}
\end{figure}

Figure~\ref{fig:noise} shows the mechanism in simulation on one ensemble of $N=40$
logistic units, every unit above the accumulation point, at the strong coupling
$\eps=0.95$. Convex weights give $\abar=3.79$ and leave the mean field chaotic
[Fig.~\ref{fig:noise}(a)], while a signed minimum-norm book with $\abar=1.80$ on the same units brings it
to rest near $\zstar$ [Fig.~\ref{fig:noise}(b)]; at this coupling the rest sits slightly off $\zstar$.
With white per-unit noise the rest persists as a tight small-noise cloud [Fig.~\ref{fig:noise}(c)]. With
fractional Gaussian per-unit noise of Hurst exponent $H$, persistent for $H>\tfrac12$ and
anti-persistent for $H<\tfrac12$, this low-pass book ($\lambda=2-\abar>0$) shows slower and
somewhat wider excursions under persistent driving ($H=0.8$) than under anti-persistent
driving ($H=0.2$) [Fig.~\ref{fig:noise}(d)].

\begin{figure}[htp]
  \centering
  \begin{subfigure}[t]{0.48\linewidth}
    \centering
    \includegraphics[width=\linewidth]{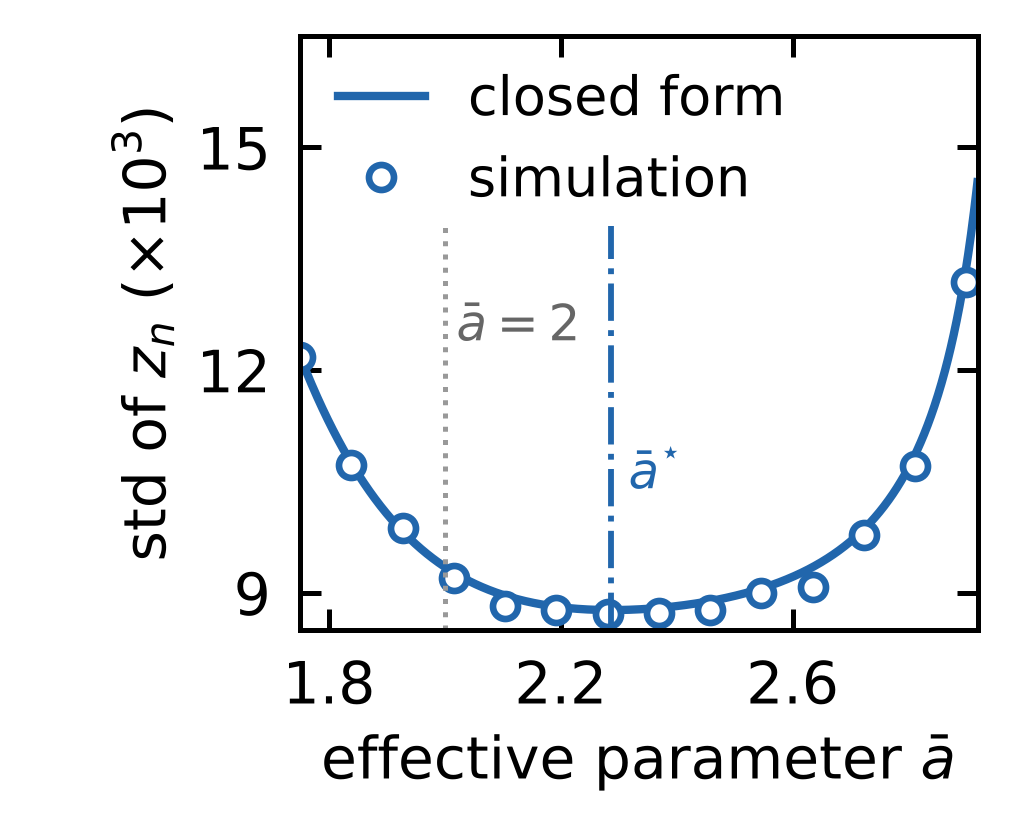}
    \caption{}
    \label{fig2a}
  \end{subfigure}
  \begin{subfigure}[t]{0.48\linewidth}
    \centering
    \includegraphics[width=\linewidth]{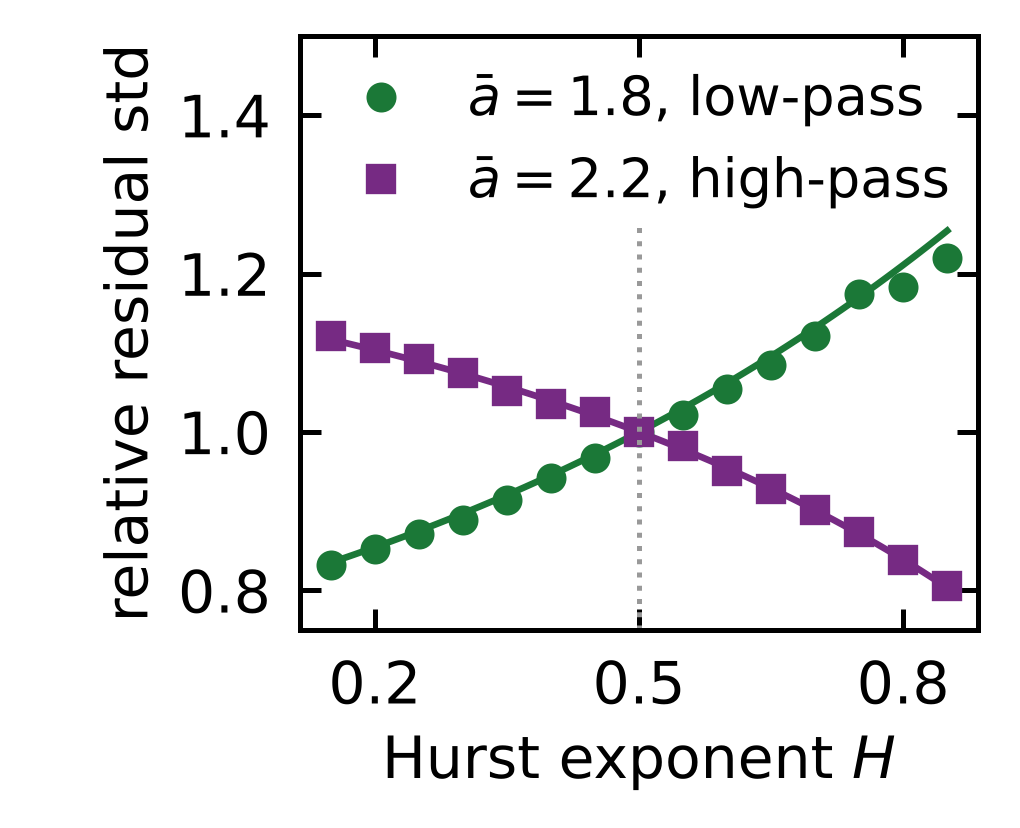}
    \caption{}
    \label{fig2b}
  \end{subfigure}
  \begin{subfigure}[t]{0.48\linewidth}
    \centering
    \includegraphics[width=\linewidth]{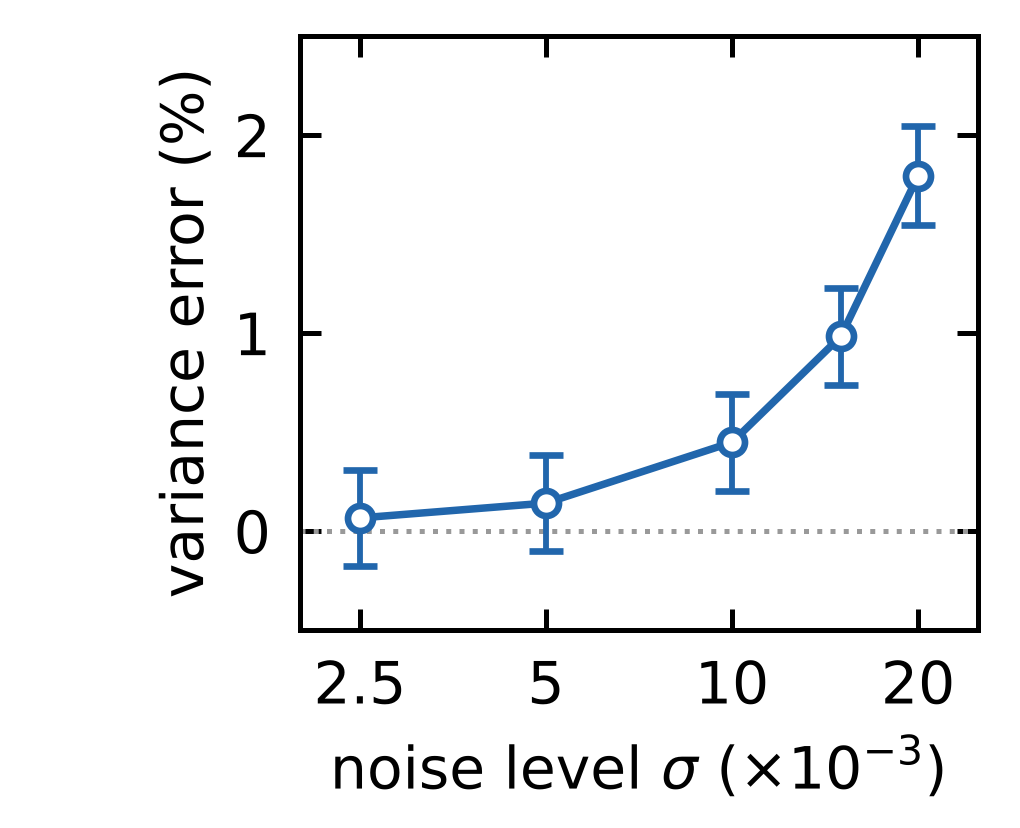}
    \caption{}
    \label{fig2c}
  \end{subfigure}
  \caption{Noise rejection at full coupling ($\eps=1$) for the $N=40$ units of
  Fig.~\ref{fig:noise}; simulation averages use initial states that reach rest. (a) White
  noise ($\sigma=0.004$): residual standard deviation of $z_n$ from the full ensemble simulation (points) and the closed form
  Eq.~\eqref{eq:varbook} (curve); the dotted line marks the flat-filter point $\abar=2$
  ($\lambda=0$) and the dash-dotted line the numerical minimizer
  $\abar^\star\!\approx\!2.29$. (b) Fractional Gaussian noise ($\sigma=0.005$): residual versus Hurst
  exponent $H$ for a low-pass design ($\abar=1.8$) and a high-pass design ($\abar=2.2$),
  each normalized to its own white level, simulation (points) and closed form
  Eq.~\eqref{eq:main} (curves); the opposite slopes cross at $H=\tfrac12$.
  (c) Convergence of the first-order form: mean signed relative error of the closed-form
  variance Eq.~\eqref{eq:varbook} against simulation at fixed pole $\abar=2.2$, simulated
  minus closed form relative to closed form, with $95\%$ confidence intervals.}
  \label{fig:noisevar}
\end{figure}

Figure~\ref{fig:noisevar}(a) evaluates the closed forms at full
coupling $\eps=1$, where the synchronization reduction is exact and the dispersion is the
demeaned previous-step noise $\delta^{(i)}_n=\xi^{(i)}_{n-1}-\bar\xi_{n-1}$. Each
$\abar$ is realized by the minimum-norm affine book used for the basin bound. The
prediction and the finite-horizon simulation agree over the tested range
$\abar\in[1.75,2.9]$, with the residual rising toward the edges.
Figure~\ref{fig:noisevar}(c) reports the mean signed relative error of the closed-form
variance against simulation: it is statistically indistinguishable from zero at the
smallest $\sigma$ and grows to about two percent by $\sigma=0.02$, consistent with the
first-order scaling and marking where the dropped terms take over. For Fig.~\ref{fig:noisevar}(c)
no initial state leaves the basin and the saturation bound is never reached, so the bounded
implementation does not bias the estimate. At the coupling $\eps=0.95$ of
Fig.~\ref{fig:noise}, finite slaving raises the floor and lowers the numerical minimizer
by about a tenth.

\emph{Choosing the book: minimum-variance weights.} The pole fixes $\abar$, but
$\abar=\sum_i w_i a_i$ and the budget $\sum_i w_i=1$ are only \emph{two} linear
constraints, $Kw=b$, with the $2\times N$ matrix $K=[\mathbf1,\mathbf a]^\top$ (rows
$\mathbf1^\top$ and $\mathbf a^\top$) and $b=(1,\abar)^\top$. These
two constraints are independent \emph{only if the units are heterogeneous}: then $\mathbf a\not\parallel
\mathbf1$, $K$ has rank two, and the budget and pole are independently assignable;
otherwise the pole constraint is degenerate (see below). The principled choice resolves
the remaining $(N\!-\!2)$-parameter freedom by minimizing the noise. At fixed pole the
white-noise variance \eqref{eq:varbook} is the quadratic form $\sigma^2 w^\top D w$ with
$D=\diag(1+\kappa a_i^2)$ and $\kappa=\lambda^2/[\abar^2(1-\lambda^2)]$, so minimizing
$w^\top D w$ over $Kw=b$ gives the \emph{unique}
\begin{equation}
  w^\star=D^{-1}K^\top\!\bigl(KD^{-1}K^\top\bigr)^{-1}b,\qquad
  w_i^\star=\frac{c_0+c_1 a_i}{1+\kappa a_i^2},
  \label{eq:wstar}
\end{equation}
where $c_0$ and $c_1$ are the two Lagrange multipliers fixed by $Kw=b$. This is
the linearly-constrained minimum-variance beamformer in the metric $D$, equivalently
the constrained Markowitz portfolio whose per-unit noise is $1+\kappa a_i^2$: the larger a
unit's parameter, the larger its parameter-weighted noise contribution under the reduction. At the flat pole $\abar=2$,
$\kappa=0$ and \eqref{eq:wstar} is the minimum-norm book $\hat w=K^\top(KK^\top)^{-1}b$,
\emph{affine in the parameter}, $\hat w_i=\alpha_0+\alpha_1 a_i$ with
$\alpha_1=(\abar-m_a)/(Nq)$, where $m_a=N^{-1}\sum_i a_i$ is the mean parameter and
$q\equiv N^{-1}\sum_i(a_i-m_a)^2$ is the \emph{parameter spread} of the list
$\mathbf a$. For a tightly spread list the metric departs little from a multiple of the
identity, and the affine book stays close to \eqref{eq:wstar}.
Since rest requires $\abar<m_a$ one has $\alpha_1<0$, so the optimum \emph{automatically}
shorts the largest-parameter units: signed weights are not imposed but are
the output of the sign-free minimum-variance problem, and adding $w_i\ge0$ makes it
\emph{infeasible} ($\abar\ge\min_i a_i>3$), so subtraction is required. The squared norm
of the affine book is $\|\hat w(\abar)\|_2^2=N^{-1}[1+(\abar-m_a)^2/q]$, least at
$\abar=m_a$ and growing as $\abar$ departs from it; a wider spread $q$ makes subtraction
cheaper. The pole is set by minimizing the variance \eqref{eq:varbook} over $\abar$: the
common-mode term carries no explicit pole gain, but through the optimized book its value
$\sigma^2\|w^\star(\abar)\|_2^2\approx\sigma^2\|\hat w(\abar)\|_2^2$ grows as $\abar$ leaves $m_a$, while the
parameter-weighted well vanishes at $\abar=2$ and diverges at the edges, so the minimizer
is interior, between $2$ and $m_a$, and is fixed by the $a_i$ (Appendix~\ref{app:opt}). With
heterogeneity the minimum-variance book at each pole is unique; the profiled outer
objective has an interior minimizer, numerically unimodal for the parameter lists tested,
though strict global uniqueness over all heterogeneous chaotic lists is not established
here. Identical units ($a_i\equiv a_1$) collapse $K$ to rank one and lock $\abar=a_1>3$,
putting rest out of reach at any leverage under output coupling: heterogeneity is
the enabling resource.

\emph{Remark (signed minimum).} Whenever the units are heterogeneous, the white-noise
rest variance has an interior minimizer over poles, numerically unimodal for the lists
tested, attained by the unique fixed-pole book. This holds \emph{even when every unit
lies above} $\ainf$: the minimizer then lies in $(2,3)$, the book \emph{necessarily}
carries negative weights, since $\abar^\star<3<\min_i a_i$ forces $\sum_i w_i a_i<\min_i
a_i$, and no non-negative book can reach it. What the construction needs is heterogeneity,
not the presence of any low-$a$ unit.

\emph{Validity of the linearization.} The exact collective dynamics is \eqref{eq:exactz};
\eqref{eq:ar1} keeps its noise forcing to first order and drops the curvature
$-\abar\,u_n^2$, the quadratic noise term $Q_n$, and the products $u_n S_n$, all of higher
order in the noise. Their effect is clearest in the field itself. At full coupling every
unit receives the noise-free common value $\bar g_n$, so $x^{(i)}_{n+1}=\bar g_n+\xi^{(i)}_n$
and $z_{n+1}=\bar g_n+\bar\xi_n$. With the parameter-weighted noise $A_n$ defined above and
$R_n=\sum_i w_i a_i\bigl(\xi^{(i)}_n\bigr)^2$, expanding each $g_i$ about $\bar g_n$ gives
exactly
\begin{equation*}
  \bar g_{n+1}-\zstar=(\lambda-2A_n)(\bar g_n-\zstar)-\abar(\bar g_n-\zstar)^2
  +\frac{\lambda}{\abar}A_n-R_n ,
\end{equation*}
so the pole of the field is the random quantity $\lambda-2A_n$. For white unit noise the
pair $(\lambda-2A_n,\,\lambda A_n/\abar-R_n)$ uses only the noise of step $n$, which makes
it independent of $\bar g_n$ and from step to step. The second moment of the linear part
then obeys a linear recursion whose coefficient is the mean-square pole
$\lambda^2+4\sigma^2\sum_i w_i^2a_i^2$, so the closed-form variances hold only while this
quantity stays below one. The condition trims the rest window slightly at each edge, and
beyond it the stationary second moment is unbounded. Near the middle of the window the
leading neglected effect is instead a noise-induced shift of the rest level. Taking expectations
of \eqref{eq:exactz} at stationarity, the curvature together with $\langle(1-2z_n)S_n
\rangle$ and $\langle Q_n\rangle$ all enter at order $\sigma^2$, so the aggregate settles
slightly off $\zstar$ by an amount set by the noise variance, a shift also found in the
noisy mean-field map of globally coupled chaotic units~\cite{demonte2004}; a closed form
for the shift here would carry $S_n$ and $Q_n$ through the average and is not attempted. The same curvature bounds robustness, because the rest has a finite basin. The simulations shown here apply a saturation
$x\mapsto\min\{1.5,\max\{-0.5,x\}\}$ at every step, so they test a \emph{bounded
implementation} of the model, not the unclipped Gaussian-driven map conditioned on
survival. For the unclipped system the local rest is metastable: a fluctuation large
enough to leave the finite basin of $\zstar$ destroys rest, and survival-conditioned
statistics would instead require terminating each trajectory at its first exit and
reporting the surviving fraction. Within the bounded implementation the closed-form
variance describes the small-noise fluctuations of
Figs.~\ref{fig:noise}--\ref{fig:noisevar}, and a trajectory that stays in the basin never
reaches the saturation bound, which acts only on replicas that have already escaped and
are excluded from the statistics.

\paragraph{Hurst grading: the pole sign selects the rejected band.}
Take fractional Gaussian noise of Hurst exponent $H$, with
$\rho_H(m)=\tfrac12\bigl(|m+1|^{2H}-2|m|^{2H}+|m-1|^{2H}\bigr)$ and spectral density
$P_H(\omega)\sim|\omega|^{1-2H}$ as $\omega\to0$: persistent noise ($H>\tfrac12$) is
low-frequency heavy, anti-persistent noise ($H<\tfrac12$) high-frequency heavy.
The filter gain $|1-\lambda e^{-j\omega}|^{-2}=(1-2\lambda\cos\omega+\lambda^2)^{-1}$
is monotone in $\cos\omega$ and peaks where the \emph{sign} of $\lambda$ dictates:
at $\omega=0$ for $\lambda>0$ (low pass), at $\omega=\pi$ for $\lambda<0$ (high pass).
To leading order in $\lambda$ the overlap integral~\eqref{eq:spectral} gives
$G-1\simeq2\lambda\,\rho_H(1)$ with $\rho_H(1)=2^{2H-1}-1$. The $H$ dependence of the
rest variance enters through this term, and since
$\partial\rho_H(1)/\partial H=(\ln2)\,2^{2H}>0$ for all $H$, the leading-order
derivative $\partial G/\partial H\simeq2\lambda\,\partial\rho_H(1)/\partial H$ carries
the sign of $\lambda$, so, to this order,
\begin{equation}
  \operatorname{sign}\!\frac{\partial\,\Var(z_\infty)}{\partial H}
  =\operatorname{sign}(\lambda)=\operatorname{sign}(2-\abar).
  \label{eq:flip}
\end{equation}
Beyond this order the relation is a numerical observation: a scan of the
fractional Gaussian covariance on $H\in[0.15,0.85]$ and $|\lambda|\le0.6$ found the sign
of $\partial\Var/\partial H$ equal to $\operatorname{sign}(\lambda)$ throughout, as did a
wider scan on $H\in[0.02,0.98]$ and $|\lambda|\le0.99$, and no general proof for all $H$
and $|\lambda|<1$ is given;
each pole position then gives a design rule.
\begin{itemize}
\item $\abar<2$ ($\lambda>0$, monotone relaxation): low-pass; suppresses
anti-persistent jitter, is most degraded by \emph{persistent} drift.
\item $\abar>2$ ($\lambda<0$, alternating relaxation): high-pass; suppresses slow
drift, is most degraded by \emph{anti-persistent} reversals.
\item $\abar=2$ ($\lambda=0$): flat; $\Var=\sigma^2\|w\|_2^2$ independent of $H$.
\end{itemize}
The two designs have opposite Hurst sensitivities, crossing at the white case
$H=\tfrac12$, the X-pattern of Fig.~\ref{fig:noisevar}(b). A designer thus places
the pole to match the expected noise color: oscillatory damping for slowly drifting
environments, monotone damping for jittery ones. This is the operational content of
the construction: noise rejection in the ensemble is \emph{designable} and
\emph{Hurst-graded}, and the design variable is a pole that positive output coupling
cannot reach.

\paragraph{Connection to adaptive filtering.}
The same algebra is standard in detection and array processing, and the identity of
minimum-variance portfolios and beamformers is known \cite{fengpalomar2016}. The Markowitz
tangency portfolio $w\propto\Sigma^{-1}\mu$ is the colored-noise matched filter for a
``signal'' $\mu$ in noise $\Sigma$. It maximizes the output signal-to-noise ratio
$(w^\top\mu)^2/(w^\top\Sigma w)$, whose maximum $\mu^\top\Sigma^{-1}\mu$ is the squared
Sharpe ratio. Collective rest is the complementary object, which uses no signal, only
$\Sigma$, and \emph{nulls} a disturbance. The minimum-variance book
$w\propto\Sigma^{-1}\mathbf1$ is the minimum-variance distortionless-response
(MVDR/Capon) beamformer~\cite{capon1969}, and a factor-neutral book, with zero
response to a common-mode steering vector $\beta$, is the linearly-constrained
minimum-variance (LCMV) beamformer~\cite{frost1972} that places a null on $\beta$. In
this dictionary the impossibility is the elementary fact that \emph{a non-negative weight
vector cannot null a strictly positive steering vector}, since $\beta^\top w>0$ whenever
$\beta>0$, $w\ge0$ and $w\ne0$. ``Rest requires subtraction'' and ``nulling a positive
steering vector requires subtraction'' are one statement. The common mode is then
multiple-access interference and the signed book a decorrelating/MMSE multiuser
detector~\cite{verdu1998}; with the temporal pole~\eqref{eq:ar1} included, the
ensemble is a space-time filter, a spatial null on the common mode, a temporal pole
on the residual.

\paragraph{Portfolio reading.}
The same construction is, term for term, Markowitz portfolio selection. Read $a_i$ as expected returns, $C$ as the
return covariance, $\abar=\sum_i w_i a_i$ as the target return, $\sum_i w_i=1$ as full
investment, and $\min w^\top Cw$ as minimum variance; then Eq.~\eqref{eq:wstar} with $D$ replaced by $C$ is the
Merton efficient-frontier weight. For uncorrelated equal-variance returns, $C=\sigma^2I$,
the squared-norm law $\|\hat w\|_2^2=N^{-1}[1+(\abar-m_a)^2/q]$ is the frontier parabola,
with the cross-sectional dispersion $q$ of expected returns as the lever; a wider
dispersion lowers the squared norm needed to hit a given target. The impossibility carries
over, and so do its consequences for constrained books.
\emph{(i)} A purely positive (long-only) book cannot null a uniformly positive common
factor: factor neutrality requires shorting, the financial face of the impossibility.
\emph{(ii)} Constrained to $w_i\ge0$, the minimum-variance portfolio is still unique
(a convex quadratic program) \emph{when the target is feasible}, i.e.\ inside the hull
$[\min_i a_i,\max_i a_i]$; it has no closed form (the binding non-negativities are not
known a priori) and is found by the standard active-set route, solve
Eq.~\eqref{eq:wstar}, zero any negative names, and re-solve until the
Karush--Kuhn--Tucker conditions hold. Below the hull, as collective rest demands, the
long-only set is empty and no minimum exists. \emph{(iii)} Even inside the hull, and again for $C=\sigma^2I$, the
\emph{unconstrained} minimum-variance weights stay long-only only in a band about the
mean, $m_a-q/(a_+-m_a)\le\abar\le m_a+q/(m_a-a_-)$ with
$a_-=\min_i a_i$; pushing the target toward
an edge forces shorting. What does \emph{not} translate is the choice of pole: in a
portfolio the target return is exogenous, whereas here the dynamical factor
$1/(1-\lambda^2)$ selects an interior optimum $\abar^\star$, a feature with no portfolio
analog.

\paragraph{Relation to prior work.}
Several ingredients used here are individually classical. The cessation of oscillation
by coupling, including by mixed attractive/repulsive, mean-field and delay-free links, is
amplitude/oscillation death
\cite{ad_review,prasad2005,karnatak2007,resmi2011,sharma2012,repulsive2014}, and coupling
of mixed sign is known to reorganize oscillator populations \cite{hongstrogatz2011}.
Heterogeneity is known to induce order in globally coupled chaotic maps
\cite{shibata1997}, the mean field of a heterogeneous population follows the map of its
mean parameter up to a shape correction \cite{demonte2003}, and microscopic noise acts on
the mean field, including as an effective-parameter shift \cite{shibata1999,demonte2004}.
The stationary variance of a first-order filter driven by long-memory Gaussian noise is
standard time-series theory \cite{hosking1981,beran}. Although these pieces are known, the
the results here are the signed book that leaves the convex hull of the parameters and so
makes a \emph{single macroscopic pole a free design variable}, the barrier that makes its
placement in the stable window \emph{impossible for positive output coupling}, and
the noise design built on that pole, whose \emph{band is selected by the pole sign and
graded by the noise memory}. The distinction from amplitude death is one of mechanism.
Amplitude death is reached by parameter mismatch, delay, or added repulsive or mean-field
links, mostly in flows; here rest is reached by a weighted parameter average over maps, so
all synchronized units share the rest value $\zstar=1-1/\abar$, and the book sets both
that value and the linearized pole.

The positive-coupling examples below do not contradict the synchronized mean-field
Proposition. Udwadia and Raju \cite{udwadia1998} study a symmetric pair of identical
coupled maps. The difference of the two units is multiplied at each step by the local
slope times a coupling factor of modulus below one, so for suitable coupling the pair
synchronizes while the
synchronized motion, which follows the single map, stays chaotic. That ordering acts on
the difference mode, needs near-identical units, and does not scale to a designable
collective rest. The nontrivial collective behavior of globally coupled chaotic maps
\cite{kaneko1990,kaneko1992,chate1992} produces an \emph{oscillating} mean field over a chaotic
substrate with strictly positive weights, a collective rhythm rather than rest. At strong
coupling a globally coupled ensemble of identical units falls into a coherent phase that
follows the single map \cite{kaneko1990b}. An unweighted heterogeneous population follows
the map of its mean parameter up to a shape correction \cite{demonte2003}; on the
synchronized set of Eq.~\eqref{eq:model} the ensemble follows the map of the weighted
parameter $\abar$, which is where the sign of the book enters. The temporal Parrondo
``chaos$+$chaos$=$order'' effect \cite{almeida2005,review2026} alternates chaotic maps in
\emph{time}, and in continuous-time systems that are linear in the parameter, periodic
parameter switching yields the attractor of the convex-averaged parameter \cite{danca2014}, a temporal analog of the convex-hull barrier. The present
construction is the spatial, static, large-$N$ counterpart, with signed weights that leave
the hull.

The positive-weight routes to rest through the units' states, sequential updating or
random delays \cite{sinha2002,atmanspacher2005,masoller2005,marti2005}, compared in the
Remark after the Proposition, rest identical units at the single-map level $1-1/a$, which
no choice of weights moves. In the output-coupled positive mechanisms above, ordering
occurs in a difference mode or as a rhythm rather than by placing the synchronized
mean-field pole inside the period-one window; the signed book implements the required
cancellation directly.

\paragraph{Empirical realization: methods.}
\label{sec:empirical}
The affine map family is special, but the \emph{linear design layer} that makes rest
possible, a signed weight book that subtracts a common component, transfers to data,
and the impossibility has a sharp financial image. This is presented as an
\emph{illustration} of the transferable linear statement, not as empirical evidence for
the dynamical theorem: asset returns are not claimed to be logistic maps, and the test is
only that a uniformly-signed common mode cannot be removed by non-negative weights and
requires subtraction, at a leverage cost. Two universes of daily returns are studied: a tightly
coupled technology cluster (``chips,'' $N=11$: ADI, AMAT, AMD, AVGO, INTC, LRCX,
MU, NVDA, QCOM, TXN, MSFT) and a diversified blue-chip set ($N=11$: AAPL, AMZN, GE,
IBM, JNJ, JPM, KO, MSFT, NVDA, TSLA, XOM). The chips panel begins in August 2009 and the
blue-chip panel in June 2010, when the last listed name (AVGO, TSLA) begins trading. All
in-sample quantities below (principal components, loadings, the four books and the
frontier) use data before 1 January 2019. Let $r^{(i)}_t$ be the daily log return of asset $i$ and
$\Sigma=\mathrm{Cov}(r)$ the $N\times N$ sample covariance. Its leading eigenpair
$(\theta_1,v_1)$ defines the \emph{common factor}: $v_1$ (the first principal
component, PC1) is the dominant collective mode, carrying the fraction
$\theta_1/\mathrm{tr}\,\Sigma$ of the total variance ($54\%$ for the chips, $42\%$
for the blue chips). Returns enter in level, so the decomposition is of the covariance,
not the correlation, matrix; the factor amplitude is $y_t=v_1^\top r_t$. The
\emph{loading} of asset $i$ is the slope of its return regressed on the factor,
$\beta_i=\mathrm{Cov}(r^{(i)},y)/\mathrm{Var}(y)$ (the division by $\mathrm{Var}(y)$
makes it a slope, not a covariance). Because $y=v_1^\top r$ with $\Sigma v_1=\theta_1
v_1$ and $\|v_1\|=1$, one has $\mathrm{Cov}(r^{(i)},y)=(\Sigma v_1)_i=\theta_1 v_{1,i}$
and $\mathrm{Var}(y)=\theta_1$, so $\beta_i=v_{1,i}$: the loading is simply the
$i$th entry of the leading eigenvector. The condition that all $\beta_i>0$ is therefore
the same as $v_1>0$, which Perron--Frobenius guarantees whenever the pairwise correlations are
positive (a non-negative irreducible matrix has a strictly positive Perron
eigenvector). For an equity cluster sharing a common market/sector factor this holds;
here every $\beta_i>0$ (range $0.16$--$0.51$ for the chips), the analog of every unit
lying above $\ainf$ ($\beta_i>0\leftrightarrow a_i>\ainf$). A book is a weight vector $w$
with $\mathbf1^\top w=1$; its factor loading is $\beta^\top w=\sum_i w_i\beta_i$, and
$\beta^\top w=0$ means the portfolio is \emph{orthogonal to the collective mode}, it
no longer rides the common factor, the equity image of rest. The comparison uses the
following books:
\begin{itemize}\setlength\itemsep{1pt}
\item equal weight, $w_i=1/N$;
\item long-only minimum variance, $\min_w w^\top\Sigma w$ s.t.\ $\mathbf1^\top w=1$,
$w\ge0$ (a quadratic program);
\item signed minimum variance, the Markowitz global minimum-variance
portfolio~\cite{markowitz1952}, $w=\Sigma^{-1}\mathbf1/(\mathbf1^\top\Sigma^{-1}\mathbf1)$,
with shorts allowed;
\item factor-neutral signed, $\min_w w^\top\Sigma w$ s.t.\ $\mathbf1^\top w=1$ and
$\beta^\top w=0$, with closed form
$w=\Sigma^{-1}K_\beta^\top(K_\beta\Sigma^{-1}K_\beta^\top)^{-1}b_\beta$, where
$K_\beta=[\mathbf1\;\,\beta]^\top$ and $b_\beta=(1,0)^\top$.
\end{itemize}
Expected returns are never used: the books minimize variance only, so no $\mu$
estimate enters. The Markowitz solution shorts of its own accord, because
$\Sigma^{-1}$ weights each eigenmode $k$ of $\Sigma$, with eigenvalue $\theta_k$, by $1/\theta_k$ and so suppresses the high-variance
collective mode. Its gross leverage is $\|w\|_1=1.4>1$ on the chips.

\begin{figure}[t]
  \centering
  \includegraphics[width=\columnwidth]{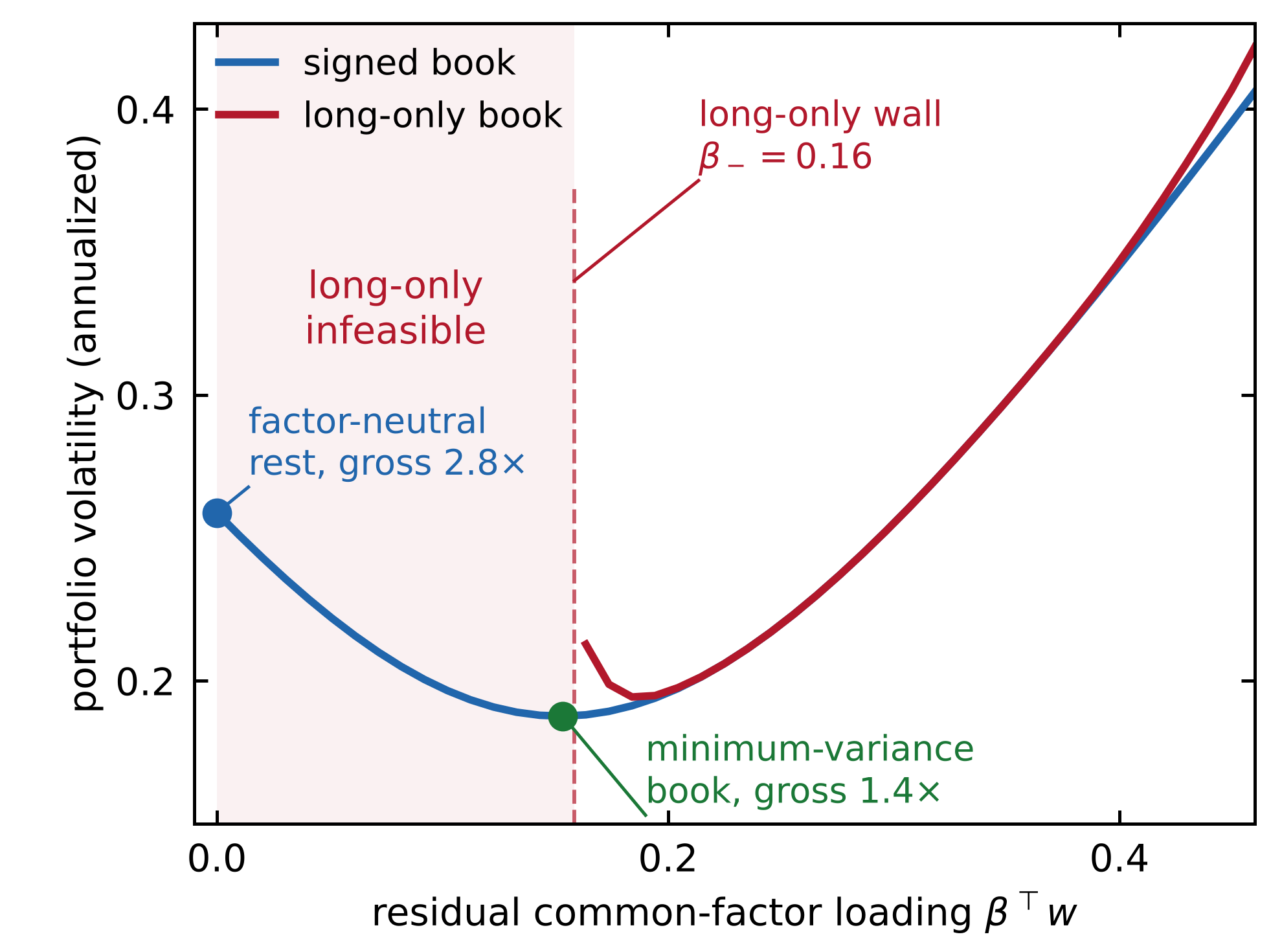}
  \caption{The impossibility and the leverage tradeoff for the chips
  (in-sample data, before 2019). Portfolio volatility versus residual common-factor loading $\beta^\top w$ (zero $=$
  factor-neutral rest). The long-only book (red) cannot cross the wall at
  $\beta_-=0.16$; the shaded region is infeasible for any non-negative book. The
  signed book (blue) continues to $\beta^\top w=0$, but reaching exact neutrality
  requires gross leverage $2.8\times$ and raises volatility; the minimum-variance book
  (gross $1.4\times$) is the lowest-risk point. The two dots mark the factor-neutral
  rest ($\beta^\top w=0$) and the minimum-variance book, each labeled with its gross
  leverage.}
  \label{fig:frontier}
\end{figure}

\begin{figure}[htp]
  \centering
  \includegraphics[width=\columnwidth]{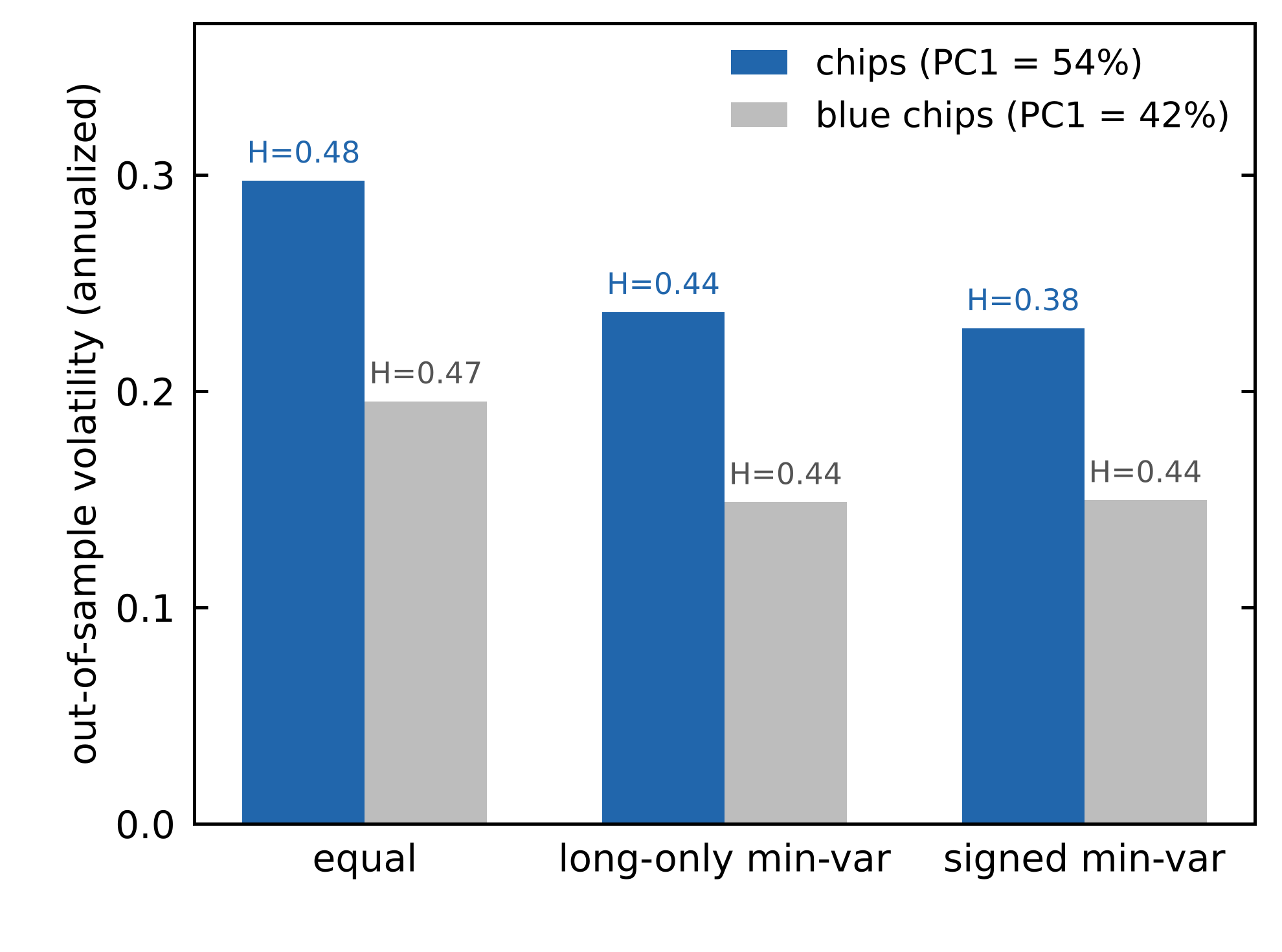}
  \caption{Walk-forward out-of-sample volatility (data, test windows 2012--2026) for three books in
  the strongly coupled chips (PC1 $=54\%$) and weakly coupled blue chips
  (PC1 $=42\%$), with the aggregate Hurst exponent $H$ annotated. The signed book gives a
  modest reduction in the chips and none in the blue chips.}
  \label{fig:oos}
\end{figure}

\paragraph{Empirical realization: results.}
The two senses of the ``wall'' must be kept distinct. For any non-negative book,
$\beta^\top w=\sum_i w_i\beta_i\ge\min_i\beta_i=:\beta_-$, with equality only when
all weight sits on the least-exposed name; here $\beta_-=0.16$. This is the
floor on the \emph{long-only} loading, a hard wall, not the loading of any one
optimized book. The achieved in-sample loadings are $\beta^\top w=0.28$ (equal), $0.19$
(long-only min-variance), $0.15$ (signed Markowitz), and $0$ (factor-neutral);
gross leverage $1.0,\,1.0,\,1.4,\,2.8$. All optimized books below
minimize the variance under the stated constraints, so the
figures show \emph{optimal} books, not arbitrary ones.
Figure~\ref{fig:frontier} plots that minimum-variance volatility as the residual
loading $\beta^\top w$ is swept: it is the empirical image of the two theoretical facts
together, the long-only wall (the impossibility) and the leverage cost of crossing it
with a signed book. Figure~\ref{fig:oos} then tests, out of sample, whether crossing it
actually pays.

\emph{(i) Impossibility.} Because all $\beta_i>0$, the long-only loading cannot fall
below $\beta_-=0.16$: the long-only factor-neutral program is infeasible. Only
a signed book attains $\beta^\top w=0$: subtraction is required, exactly as a
negative weight is required to pull $\abar$ below $\ainf$. (That the signed Markowitz
loading, $0.15$, lands near $\beta_-$ is incidental; one is a single-name
vertex, the other a diversified signed portfolio.)

\emph{(ii) Leverage is the cost, and over-subtraction re-inflates variance.}
Forcing $\beta^\top w=0$ costs gross leverage $2.8$ and \emph{raises} volatility,
whereas the signed minimum-variance book reaches its optimum at $\|w\|_1=1.4$. This
is the portfolio image of the floor $w^\top C w$ of the cross-correlated rest variance with
$C=\Sigma$: when the residual noise is uncorrelated and of equal variance across units,
$\Sigma\propto I$, one has $w^\top\Sigma w\propto\|w\|_2^2$, and at the flat pole the
minimum-variance book \emph{is} the minimum-norm book used for the basin bound, so
there least $\|w\|_2$ and least noise coincide. For a general correlated
$\Sigma$ the minimum-variance book minimizes $w^\top\Sigma w$, not the gross leverage
or $\|w\|_2$, and the two need not agree.

\emph{(iii) When subtraction helps.} The mechanism only matters if subtracting the
common mode lowers risk \emph{out of sample}, on data the weights never saw: this is
what Fig.~\ref{fig:oos} tests. In a walk-forward out-of-sample test
(three-year trailing window, quarterly rebalancing, Ledoit--Wolf shrinkage
covariance~\cite{ledoit2004}; weights formed only from past data; test windows from mid-2012
for the chips and mid-2013 for the blue chips, through 2026), the signed book
gives a modest reduction in out-of-sample volatility in the chips ($0.229$ against $0.237$
long-only and $0.297$ equal weight) and the lowest persistence (aggregate Hurst exponent
by detrended fluctuation analysis~\cite{peng1994}, $0.48\!\to\!0.38$), while in the blue
chips shorting yields essentially nothing ($0.150$ against $0.149$); Fig.~\ref{fig:oos}.
A longer chips history from 2000, with AVGO dropped and a fixed diagonal shrinkage of the
covariance, has test windows that include the 2008 crisis and shows the same ordering
($0.238$ against $0.244$). The gain is, however, not a monotone function of factor
strength. Random baskets of six to ten names were drawn from the stocks above and
five sector exchange-traded funds (XLB, XLE, XLF, XLI, IYR), with the books estimated
before 2019 from the sample covariance, without shrinkage, and evaluated afterwards. Across them the correlation between the
common-factor share and the out-of-sample gain is weak, and the basket of the five sector
funds, with a very strong common factor (PC1 $=79\%$), gains nothing. A strong common factor is therefore not sufficient for a gain, and
subtraction helps only when a strong shared mode coexists with substantial
mode-orthogonal (idiosyncratic) variance to redeploy, as in a single-name cluster;
cancel the mode in an already-diversified basket and only leverage noise remains. The persistence shift is reported as supporting, not central: daily returns are
near-martingale ($H\approx0.5$), so the robust, model-independent findings are the
impossibility and the leverage/variance tradeoff.

\paragraph{Conclusion.}
Collective rest produced by signed coupling realizes a designable one-pole
linearization. Equations~\eqref{eq:main}--\eqref{eq:varbook} give its leading-order
small-noise variance at full coupling, from the pole $\lambda=2-\abar$, the plain and
parameter-weighted norms of the book, and the noise-color factor $G$, while Eq.~\eqref{eq:flip} gives the Hurst-sign
relation to leading order in $\lambda$ and is numerically supported beyond that order on
the tested grid. For every fixed pole the minimum-variance book is unique under
heterogeneity; the profiled pole has an interior numerical minimizer for the parameter
lists studied, and global uniqueness is not claimed. For the specified sequence of
diversified books, the guaranteed basin fraction approaches one with $N$ under the stated
randomized-initial-condition assumptions, at a rate set by the parameter spread $q$
[Eq.~\eqref{eq:hetrate}], so the same heterogeneity that makes the pole placeable also
raises the guaranteed basin. Heterogeneity is not incidental but the enabling resource:
it is pole-controllability (the parameter vector adds a steering direction
independent of the budget), the squared norm in excess of $1/N$ needed to place the pole
scales as $1/q$ in the spread, and under output coupling identical units lock the pole outside the rest
window so that no book, signed or not, can rest. The capability rests on a barrier:
positive output coupling cannot place the pole, so under output coupling
Hurst-matched rejection requires subtraction over \emph{heterogeneous} units. The mechanism is not tied
to the logistic map: its algebra extends to any affine (parameter-scaling) family
$g_a=a\,h(x)$, with the family-specific concentration constants recomputed, and it is
supported numerically in the non-affine Ricker family, where no single effective
parameter exists. Natural further extensions are
continuous-time ensembles, where the pole becomes a contraction rate. A higher-order
route is memory or delay in the coupling,
$z_{n+1}=\mathcal F(z_n,\dots,z_{n-p+1})+\bar\xi_n$, which linearizes to an order-$p$
recursion $u_{n+1}=\sum_{k=1}^{p}\phi_k u_{n+1-k}+\bar\xi_n$ with coefficients $\phi_k$ and $p$ poles, the roots of
the companion characteristic polynomial. The book then places \emph{several} poles,
complex-conjugate pairs included, turning the rest state into a full IIR designer:
notch or band-reject responses matched to a structured noise spectrum, with the
present AR(1) the minimal memoryless case (Appendix~\ref{app:ar2}). Delay coupling ties this to the delay route
to amplitude death, and there convexity no longer excludes rest, since positive weights
with random delays settle identical chaotic maps to the single-map fixed point
\cite{masoller2005,marti2005}. Coupling through the units' states and sequential updating
do the same \cite{sinha2002,atmanspacher2005}. Whether these positive couplings can also set
the rest level and its pole by design is open.

\section*{Funding and Competing Interests}
The author received no external funding for this work and declares no competing
interest.

\section*{Data Availability}
The daily price series underlying the empirical results and the corresponding figures
were obtained from Stooq (\url{https://stooq.com}), which provides freely available
historical market data, and no proprietary data were used.

\section*{Acknowledgment}
AI tools were used mainly for editorial assistance including proofreading, literature search
and language refinement. All theoretical developments, derivations and conclusions are the
author's work.

\appendix
\section{Rest below full coupling}
\label{app:below}

Write the noise-free Eq.~\eqref{eq:model} as $\mathbf x_{n+1}=F(\mathbf x_n;\eps)$, with
$\mathbf x=[x^{(1)},\dots,x^{(N)}]^\top$. Its Jacobian is the $N\times N$ matrix
\begin{equation}
\begin{aligned}
  J&=M \Gamma,\qquad M=(1-\eps)I+\eps\,\mathbf 1\,w^\top,\\
  \Gamma&=\diag\bigl[g_1'(x^{(1)}),\dots,g_N'(x^{(N)})\bigr],
\end{aligned}
  \label{eq:jacN}
\end{equation}
where $\mathbf 1$ is the $N$-vector of ones and $g_i'(x)=a_i(1-2x)$. Fix a closed bounded
box $B$ that contains the unit cube, such as the saturation box of the simulations. A rest
is a fixed point of $F$ in $B$ at which the spectral radius of $J$ is below one. The map
$F$ is polynomial in $\mathbf x$ and affine in $\eps$ and in the weights, so the residual
$\mathbf x-F(\mathbf x;\eps)$ and the spectral radius of $J$ are continuous in all of these quantities.
Since $\sum_i w_i=1$, the matrix $M$ has eigenvalue one along $\mathbf 1$ and eigenvalue
$1-\eps$ on the $(N-1)$-dimensional subspace orthogonal to $w$, so at every point
\begin{equation}
  \det J=(1-\eps)^{N-1}\prod_{i=1}^{N} g_i'\bigl(x^{(i)}\bigr).
  \label{eq:detN}
\end{equation}
A rest therefore requires $\prod_{i=1}^{N}|g_i'(x^{(i)})|<(1-\eps)^{-(N-1)}$, a bound on the
unit slopes at the fixed point that tightens toward one as the coupling weakens.

\emph{Exclusion lemma:} Let the weights range over a compact set $W$. Suppose that at a
coupling $\eps_0$ every fixed point of $F(\,\cdot\,;\eps_0)$ in $B$ has spectral radius of
$J$ above one for every $w\in W$. Then there is a $\tau>0$ such that for
$|\eps-\eps_0|<\tau$ no fixed point in $B$ has spectral radius at most one, for any
$w\in W$. If this failed, there would be couplings $\eps_\ell\to\eps_0$, books $\tilde w_\ell\in W$
and fixed points $\mathbf y_\ell\in B$ with spectral radius at most one. Compactness of
$B\times W$ gives a subsequence converging to some $(\mathbf x,w)$, continuity of the
residual makes $\mathbf x$ a fixed point at $\eps_0$, and continuity of the spectral radius
keeps it at most one there, against the hypothesis. The same argument holds with the
parameters $a_i$ also ranging over a compact set, which makes $\tau$ uniform in them.

\emph{No rest at weak coupling:} At $\eps=0$ the units are uncoupled, and a fixed point has each
$x^{(i)}$ equal to $0$ or $1-1/a_i$, and $J=\Gamma$ has diagonal entries $a_i$ or $2-a_i$.
Every unit lies above $\ainf>3$, so each entry has modulus above one and every fixed point
has spectral radius above one. The lemma at $\eps_0=0$, with $W$ any closed bounded set of
books, signed or not, excludes rest for all couplings below some $\tau>0$. Here the exclusion is not confined to
$B$. For $\eps<1$ the quadratic part of the fixed-point equations,
$(1-\eps)a_i\bigl(x^{(i)}\bigr)^2+\eps\sum_k w_k a_k\bigl(x^{(k)}\bigr)^2=0$ for every $i$,
gives $\sum_k w_k a_k\bigl(x^{(k)}\bigr)^2=0$ on multiplying by $w_i$ and summing, and then
each $x^{(i)}=0$. A sequence of fixed points leaving every bounded set, rescaled to unit
length, would converge to a nonzero root of this quadratic part, so for couplings in
$[0,\tfrac12]$ and books and parameters in compact sets the fixed points stay in a bounded
set, and $B$ can be taken to contain them all.

\emph{No rest under positive weights near full coupling:} At $\eps=1$ every unit receives the
field, $F(\mathbf x;1)=\bar g\,\mathbf 1$, so every fixed point is $\mathbf x=s\mathbf 1$ with
$s=\abar s(1-s)$, that is $s=0$ or $s=\zstar$. There $J=\mathbf 1\,w^\top \Gamma$ has rank
one, with $N-1$ zero eigenvalues and the eigenvalue $w^\top \Gamma\,\mathbf 1=\abar(1-2s)$, equal to
$\abar$ at the origin and $2-\abar$ at $\zstar$. For a convex book the Proposition gives
$\abar>\ainf>3$, so this eigenvalue has modulus above one at both fixed points. The lemma at
$\eps_0=1$, with $W$ the simplex of convex books and the $a_i$ in any compact subset of
$(\ainf,4]$, excludes rest for all couplings in some window $(1-\tau,1]$, uniformly in the
book and the parameters.

\emph{Signed rest near full coupling:} Let a signed book place $\abar$ in $(1,3)$. At
$\eps=1$ the point $\zstar\mathbf 1$ is a fixed point of $F$, and $J$ there has eigenvalues
$0$, repeated $N-1$ times, and $2-\abar$, all inside the unit circle. The derivative of the
residual with respect to $\mathbf x$ is $I-J$, with eigenvalues $1$ and $\abar-1$, so it is
invertible. The implicit function theorem gives a fixed point $\mathbf x^\ast(\eps)$ that
depends continuously on $\eps$ near one and equals $\zstar\mathbf 1$ at $\eps=1$. The
spectral radius of $J$ at $\mathbf x^\ast(\eps)$ is continuous in $\eps$ and equals
$|2-\abar|<1$ at $\eps=1$, so it stays below one on some window $(1-\tau,1]$, on which
$\mathbf x^\ast(\eps)$ is a locally attracting rest. The window cannot reach $\eps=0$, where
the weak-coupling exclusion applies.

Between these windows the absence of rest under positive weights is not proved. For two
units every fixed point is a real root of a quartic, and solving for all of them on a fine
grid of couplings, of convex splits of the book and of both parameters over the chaotic
range finds none, inside or outside $B$, with spectral radius at most one.

\section{Choosing the optimal book: nested minimization and the optimal pole}
\label{app:opt}
A book enters the rest filter only through two scalars, the effective parameter
$\abar=\sum_i w_i a_i$, which sets the pole $\lambda=2-\abar$, and the residual noise, so the variance-optimal book is found in two nested steps.
Write $m_a=N^{-1}\sum_i a_i$ for the mean parameter and
$q\equiv N^{-1}\sum_i(a_i-m_a)^2$ for the (deterministic) spread of the list
$\mathbf a=(a_1,\dots,a_N)$.

\emph{Inner step: weights at a fixed pole.} Fix $\abar$ and minimize the rest variance
over $w$. For independent units this is $\min_w w^\top D w$ with $D=\diag(1+\kappa a_i^2)$,
$\kappa=\lambda^2/[\abar^2(1-\lambda^2)]$, subject to $\mathbf1^\top w=1$ and
$\mathbf a^\top w=\abar$, solved by the beamformer Eq.~\eqref{eq:wstar}. At the flat pole
$\kappa=0$, $D=I$, and the solution is the minimum-norm book
$\hat w=K^\top(KK^\top)^{-1}b$; for a clustered list $D$ departs little from a multiple
of the identity, so this affine book is the tight approximation used here. Because
$\hat w$ lies in the row space of $K$ it is \emph{affine in the parameter},
$\hat w_i=\alpha_0+\alpha_1 a_i$; the two coefficients follow from the two constraints,
\begin{equation}
  \alpha_0+\alpha_1 m_a=\tfrac1N,\qquad \alpha_0 m_a+\alpha_1(q+m_a^2)=\tfrac{\abar}{N},
\end{equation}
whose solution ($m_a\times$ the first subtracted from the second) is
$\alpha_1=(\abar-m_a)/(Nq)$, $\alpha_0=\tfrac1N-\alpha_1 m_a$, i.e.
\begin{equation}
  \hat w_i=\frac1N+\frac{(\abar-m_a)(a_i-m_a)}{Nq}.
  \label{eq:waffine}
\end{equation}
This is self-consistent rather than circular: $\abar$ enters as the chosen target and
the book reproduces it, $\sum_i \hat w_i a_i=m_a+(\abar-m_a)\,N^{-1}q^{-1}\sum_i
a_i(a_i-m_a)=m_a+(\abar-m_a)=\abar$, since $\sum_i a_i(a_i-m_a)=Nq$. \emph{Adding the
sign constraint $w_i\ge0$ makes the program infeasible} for rest: a non-negative book
has $\abar\ge\min_i a_i>3$ and cannot meet $\abar<3$, so shorting is a feasibility
requirement, not a stylistic one, and \eqref{eq:waffine} supplies it automatically,
$\alpha_1<0$ shorting the largest-$a_i$ units. More generally \eqref{eq:waffine} stays
non-negative only in a band about the mean,
\begin{equation}
  m_a-\frac{q}{a_+-m_a}\;\le\;\abar\;\le\;m_a+\frac{q}{m_a-a_-},
  \label{eq:posband}
\end{equation}
so the unconstrained minimum-norm book, the minimum-variance book when $D\propto I$, is
long-only only for targets near $m_a$; any rest pole, lying below the whole cluster, falls
outside it. The squared norm of \eqref{eq:waffine} follows from
$\|\hat w\|_2^2=b^\top(KK^\top)^{-1}b$ with
$KK^\top=\bigl(\begin{smallmatrix}N & Nm_a\\ Nm_a & N(q+m_a^2)\end{smallmatrix}\bigr)$,
whose determinant is $N^2q$; expanding gives
\begin{equation}
  \|\hat w(\abar)\|_2^2=\frac1N\Bigl[1+\frac{(\abar-m_a)^2}{q}\Bigr].
  \label{eq:lev}
\end{equation}
For cross-correlated (temporally white) unit noise of covariance $C$ the same
construction applies with $D=C+\kappa\,\diag(\mathbf a)\,C\,\diag(\mathbf a)$ in place of
$\diag(1+\kappa a_i^2)$, and the constrained minimizer is again Eq.~\eqref{eq:wstar} with
this $D$ (the two agree when $C=\sigma^2 I$). Replacing $\|w\|_2^2$ by $w^\top C w$ alone
omits the parameter-weighted covariance $\diag(\mathbf a)\,C\,\diag(\mathbf a)$.

\emph{Outer step: the pole.} At the minimum-variance book the white-noise variance
Eq.~\eqref{eq:varbook} is a function of the pole alone,
$V(\abar)=\sigma^2\,b^\top(KD^{-1}K^\top)^{-1}b$, with $D=\diag(1+\kappa a_i^2)$ and
$\kappa=\lambda^2/[\abar^2(1-\lambda^2)]$. At fixed weights the common-mode contribution
has unit transfer gain; after profiling it varies through $\|w^\star(\abar)\|_2^2$, close to
$\|\hat w(\abar)\|_2^2$ and growing
as $\abar$ leaves $m_a$, while the parameter-weighted part vanishes at $\abar=2$
($\kappa=0$) and diverges at the edges $\abar\to1,3$ ($\kappa\to\infty$), so $V$ has an
interior minimizer. It is not the root of
a quadratic, since $\kappa$ enters $D$ nonlinearly; it is obtained numerically and is fixed
by the list $\mathbf a$ through $D$. It lies between $2$ and $m_a$: the
parameter-weighted well pulls the pole toward $\abar=2$, the squared norm pulls it toward
$m_a$, and the balance sits between them. For the ensemble of Fig.~\ref{fig:noisevar} the
minimizer is $\abar^\star\approx2.29$, marked in Fig.~\ref{fig:noisevar}(a). The affine
book used there gives the same minimizer to this precision, and the simulated floor is
minimized at the same pole. For colored noise the factor
$G(\lambda;\rho)$ re-weights frequencies and moves the optimum, toward $\abar=2$ for
anti-persistent noise, into $(1,2)$ only for strong anti-persistence, and toward $3$ for
persistent noise, so the general bracket is $(1,3)$.

\emph{Heterogeneity is the resource.} Equation~\eqref{eq:lev} shows that the squared norm in
excess of $1/N$ scales as $1/q$: the spread of the $a_i$ is what makes pole placement affordable. The
degenerate case is decisive: for \emph{identical} units $a_i\equiv a_1$, one has
$\abar=a_1\sum_i w_i=a_1$ \emph{locked} for every book, signed or not, so no weighting moves
the pole off $a_1>3$ and rest is unreachable at any leverage. Under output
coupling, designable rest needs \emph{both} signed weights (to escape the convex hull)
\emph{and} heterogeneity $q>0$ (to have a hull to escape), at an excess squared-norm cost
$\propto1/q$.

\emph{Why one pass is global.} Because $\abar=\sum_i w_i a_i$ is itself a function of
$w$, the two steps are not minimization over independent blocks but exact
\emph{profiling}: with $\mathcal V(w)$ the white-noise rest variance of book $w$,
$\min_w \mathcal V(w)=\min_{\abar}\min_{w:\,\mathbf a^\top w=\abar}\mathcal V(w)$, an
identity valid regardless of convexity. The inner minimization is a convex quadratic
program (unique global $w^\star(\abar)$, closed form); the outer is one-dimensional
(the interior minimizer of $V(\abar)$ on $(1,3)$, found directly). The sequential solve
therefore returns the \emph{global} optimum of the joint
problem in a single pass, even though that objective is not jointly convex in $w$. This
is \emph{not} alternating projection: iterating, recomputing $w^\star(\abar^\star)$ and
re-optimizing the pole, reproduces the same pair, since the profiled outer objective
does not depend on the current $w$. The fixed point is reached at once, and it is global.

\section{Explicit higher-order design: a two-pole rest filter}
\label{app:ar2}
Give the mean-field coupling a one-step propagation delay, so the map acts on a
blend of the present and previous collective coordinate,
\begin{equation}
  z_{n+1}=\abar\,\zeta_n(1-\zeta_n)+\bar\xi_n,\qquad
  \zeta_n=(1-\eta)z_n+\eta z_{n-1},
\end{equation}
with delay weight $\eta\in[0,1]$ ($\eta=0$ recovers the memoryless case). The fixed point is
unchanged; with $u_n=z_n-\zstar$ and $\lambda=2-\abar$, linearization gives the
second-order recursion
\begin{equation}
  u_{n+1}=\lambda(1-\eta)\,u_n+\lambda\eta\,u_{n-1}+\bar\xi_n\,,
\end{equation}
an AR(2) with coefficients $\phi_1=\lambda(1-\eta)$, $\phi_2=\lambda\eta$ and transfer
function $\Phi(\nu)=1/(1-\phi_1\nu^{-1}-\phi_2\nu^{-2})$. The two poles of this filter are
$\nu_{1,2}=\tfrac12\big(\phi_1\pm\sqrt{\phi_1^2+4\phi_2}\big)$. The decisive freedom is that
they may be \emph{complex}: since $\phi_1^2+4\phi_2=\lambda[\lambda(1-\eta)^2+4\eta]$ and
$\lambda=2-\abar<0$ when $\abar>2$, the discriminant is negative precisely when
$(\abar-2)(1-\eta)^2<4\eta$ (with $\abar>2$, $\eta>0$), and the conditions $\abar>2$
and $\eta>0$ are necessary but not by themselves sufficient. In that range the pole
pair is $|\nu_1|e^{\pm j\omega_0}$ with $|\nu_1|=\sqrt{(\abar-2)\eta}$ and
$\cos\omega_0=\phi_1/(2|\nu_1|)$, a resonance whose center $\omega_0$ and sharpness $|\nu_1|$ the
book and delay weight tune. For white driving the stationary variance is the standard
AR(2) result
\begin{equation}
  \Var(z_\infty)=\sigma^2\|w\|_2^2\,
  \frac{1-\phi_2}{(1+\phi_2)\big[(1-\phi_2)^2-\phi_1^2\big]} .
\end{equation}
This appendix is a \emph{reduced-order design extension}: the delayed scalar recursion
above is posited directly, not derived from a delayed microscopic ensemble, and the
additive $\bar\xi$ forcing omits the noise-dispersion corrections ($S_n$, $Q_n$) that
were essential in the AR(1) analysis. The variance above is therefore exact for the
stated abstract AR(2) recursion, not automatically for the underlying noisy ensemble;
what the appendix establishes is that placing several poles, complex pairs
included, is available once the coupling carries memory.
A $p$-tap profile $\{\eta_k\}_{k=0}^{p-1}$ ($\sum_k\eta_k=1$) generalizes this to
$u_{n+1}=\lambda\sum_k\eta_k u_{n-k}+\bar\xi_n$, an all-pole filter
$\Phi(\nu)=1/(1-\lambda\sum_k\eta_k \nu^{-(k+1)})$ whose $p$ poles are the roots of
$\nu^p=\lambda\sum_k\eta_k \nu^{p-1-k}$: the book sets the gain $\lambda=2-\abar$, the
delay profile shapes where the poles sit.

\begin{figure}[htp]
  \centering
  \begin{subfigure}[t]{0.48\linewidth}
    \centering
    \includegraphics[width=\linewidth]{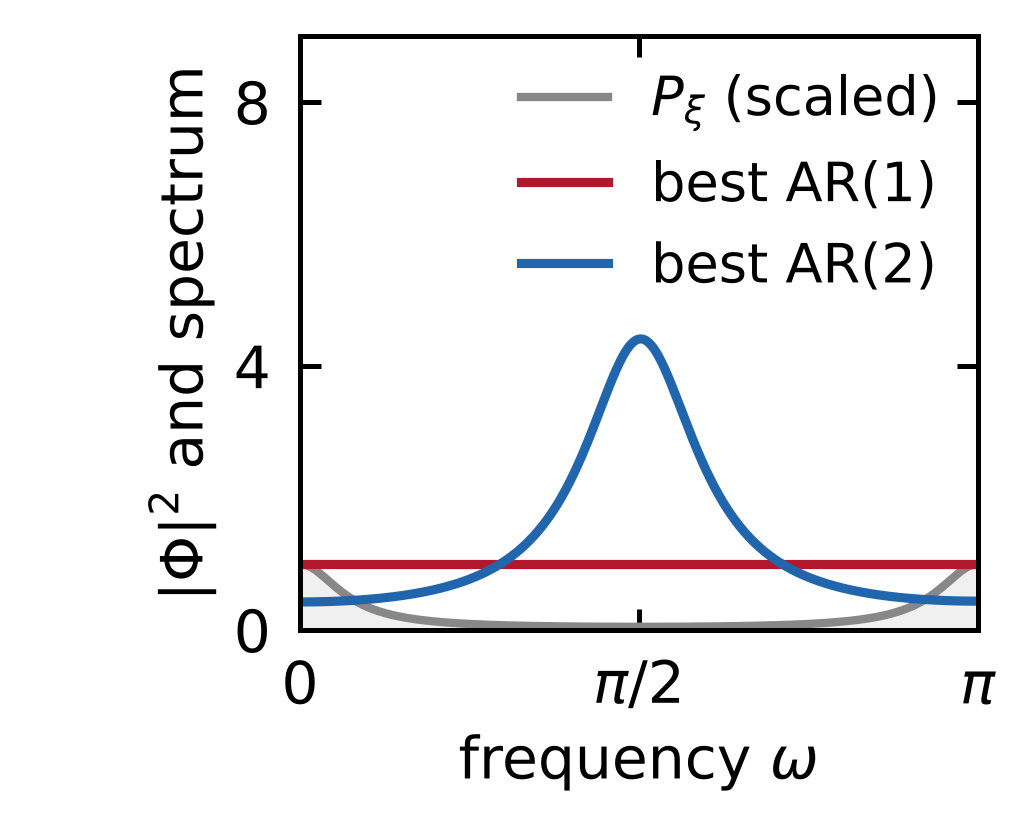}
    \caption{}
    \label{fig5a}
  \end{subfigure}
  \hfill
  \begin{subfigure}[t]{0.48\linewidth}
    \centering
    \includegraphics[width=\linewidth]{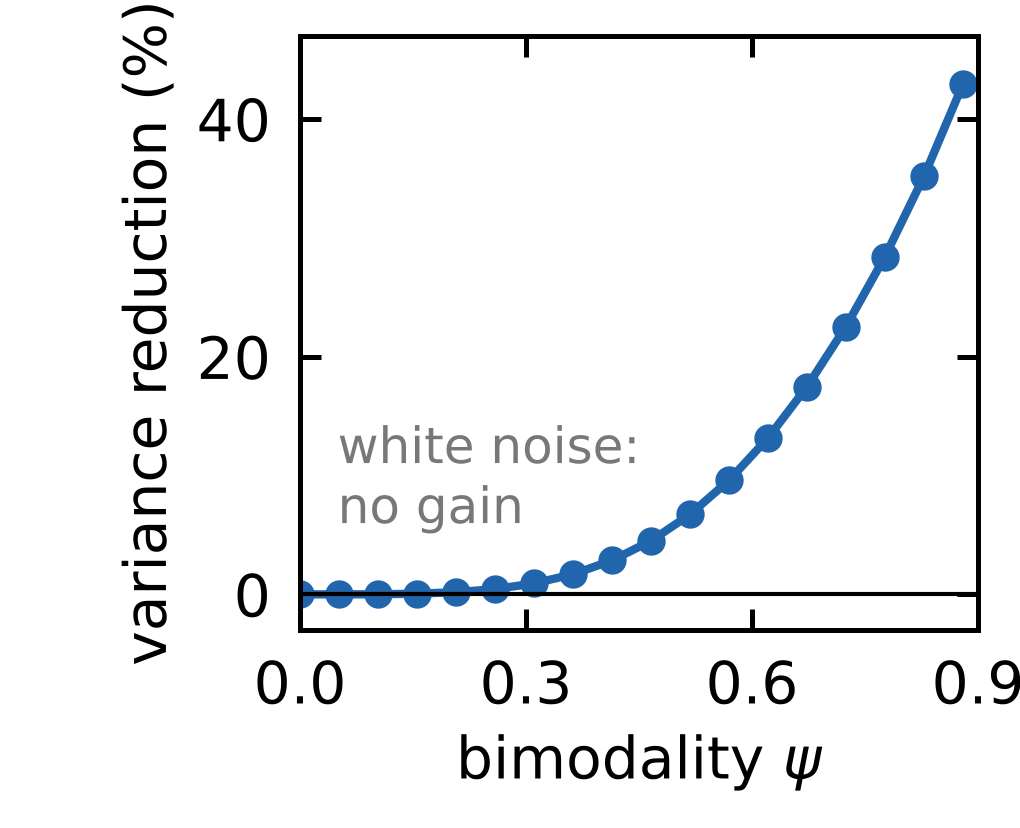}
    \caption{}
    \label{fig5b}
  \end{subfigure}
  \caption{Two poles beat one only for colored noise (closed-form responses). (a) For
  bimodal noise with $\psi=0.8$ (spectrum $P_\xi$, gray; energy at $\omega=0$ and $\pi$), the best AR(1)
  response is flat
  ($\lambda=0$, red): one real pole cannot reject both bands. The best AR(2) response
  (blue) is small at both bands and resonant in between, attenuating the noise where it
  lives. (b) Rest-variance reduction of the best AR(2) over the best AR(1) versus
  the bimodality $\psi$ ($\psi=0$ white): zero for white noise, rising to about $43\%$
  at the largest $\psi$ shown as the noise becomes strongly bimodal. The book norm
  $\|w\|_2$ is held fixed across designs.}
  \label{fig:arp}
\end{figure}

Whether higher order \emph{helps} depends entirely on the noise. For white noise the
single pole at the origin ($\abar=2$) is optimal at fixed book norm $\|w\|_2$, and extra
poles can only add variance, so AR($p$) gives nothing [Fig.~\ref{fig:arp}(b), white noise].
For structured noise a complex pole pair rejects bands a single real pole cannot.
Symmetric bimodal noise, with spectrum
$P_\xi(\omega)=(1-2\psi\cos\omega+\psi^2)^{-1}+(1+2\psi\cos\omega+\psi^2)^{-1}$ for
$0\le\psi<1$, is white at $\psi=0$ and concentrates its energy at both $\omega=0$ and
$\pi$ as $\psi\to1$. It defeats AR(1) outright, since one real pole can attenuate only one
band and the AR(1) optimum is the do-nothing $\lambda=0$, while the AR(2) resonance
suppresses both bands and cuts the rest variance by about $43\%$ at the largest $\psi$
shown, rising as $\psi\to1$ (Fig.~\ref{fig:arp}). AR($p$) is therefore superior only
conditionally: equal to AR(1) for
white noise, increasingly better as the noise departs from white, at the cost of the
delay structure that the extra poles require.

\medskip
\noindent\emph{Remark (the dollar-neutral limit $\sum_i w_i=0$).}
Relaxing the budget from $\sum_i w_i=1$ to $\sum_i w_i=0$ changes the character of the
construction entirely. The collective coordinate $z_n=\sum_i w_i x_n^{(i)}$ is then a
\emph{contrast}, blind to any common level: writing $x_n^{(i)}=X_n+d_n^{(i)}$ gives
$z_n=X_n\sum_i w_i+\sum_i w_i d_n^{(i)}=\sum_i w_i d_n^{(i)}$, perfect common-mode
rejection. The $\varepsilon$-cancellation is incomplete, $z_{n+1}=(1-\varepsilon)\bar
g_n+\bar\xi_n$, and $\bar a=\sum_i w_i a_i$ is now an unconstrained contrast rather than
a hull-bounded average, so the period-doubling impossibility no longer applies, but
the price is that, for identical units, the contrast can only rest at $z=0$ (synchronization), never at a
designable level. For identical units ($a_i\equiv a_1$), linearizing about the noise-free synchronized orbit $s_n$, which obeys
$s_{n+1}=(1-\varepsilon)g(s_n)$ since the field vanishes in sync, gives
$z_{n+1}=(1-\varepsilon)g'(s_n)\,z_n$. The transverse Lyapunov exponent therefore equals
the exponent $\Lambda_\parallel$ of the synchronized orbit itself,
$\Lambda_\perp=\ln(1-\varepsilon)+\langle\ln|g'|\rangle=\Lambda_\parallel$, and $z=0$ is
stable exactly when this exponent is negative,
$\Lambda_\perp=\ln(1-\varepsilon)+\langle\ln|g'|\rangle<0$; this is the criterion, not any
statement about a parameter threshold, since the renormalized map
$(1-\varepsilon)a_1\,x(1-x)$ can carry stable periodic windows above $3$ that are equally
non-chaotic. A simple \emph{sufficient} condition is that the renormalized map have a
stable fixed point, $(1-\varepsilon)a_1<3$ (i.e.\ $\varepsilon>1-3/a_1$), when strong
coupling drags the effective parameter into the fixed-point window and stabilizes the
null; the full non-chaotic range is larger. For heterogeneous units, genuine
rest of the contrast requires the \emph{double null} $w\perp\{\mathbf1,\mathbf a\}$
($\sum_i w_i=0$ and $\bar a=0$), the LCMV condition of the adaptive-filtering
dictionary, after which the transverse dynamics form a moment hierarchy that does not
close at scalar order; its stability is left to future work.

\end{document}